\documentclass[12pt]{article}
\usepackage{amssymb,amsmath,amscd}
\usepackage{epsfig}
\usepackage{epstopdf}
\usepackage{latexsym}
\usepackage{graphicx}
\usepackage{booktabs}
\usepackage{bbm}
\usepackage[margin=15pt,small]{caption}
\usepackage{subcaption}
\usepackage{enumitem}
\usepackage{float}
\usepackage[toc]{appendix}
\usepackage[numbers,compress]{natbib}
\usepackage{tikz}
\usetikzlibrary{positioning}
\usepackage{array,longtable}

\usepackage[T1]{fontenc}

\usepackage{color}
\usepackage{datetime}

\usepackage[
      colorlinks=false,
      linkcolor=darkblue,  
      urlcolor=blue,    
      filecolor=blue,     
      citecolor=red,
linktocpage=true,
      pdfstartview=FitV,
      bookmarksopen=true    
      ]{hyperref}

\setlist{nolistsep}

\newcolumntype{L}{>{$}l<{$}}
\allowdisplaybreaks

\ifpdf
\fi

\let\oldbibliography\thebibliography
\renewcommand{\thebibliography}[1]{\oldbibliography{#1}
\setlength{\itemsep}{0pt}} %Reducing spacing in the bibliography.

\newcommand*{\boxedcolor}{red}
\makeatletter
\renewcommand{\boxed}[1]{\textcolor{\boxedcolor}{%
  \fbox{\normalcolor\m@th$\displaystyle#1$}}}
\makeatother

\definecolor{cardinal}{rgb}{0.6,0,0}
\definecolor{darkgreen}{rgb}{0,0.5,0}
\definecolor{golden}{rgb}{0.92, 0.7, 0}
\definecolor{midnight}{rgb}{0, 0, 0.5}
\definecolor{darkblue}{rgb}{0.2, 0, 0.8}

\def\Re{{\rm Re}} \def\Im{{\rm Im}}

\newcommand{\citeurl}[1]{\href{#1}{\small\texttt{#1}}}

\def\coeff#1#2{\relax{\textstyle {#1 \over #2}}\displaystyle}
\def\ds{\displaystyle}

\def\ZZ{\mathbb{Z}}

\def\eql{~=~}

\def\cals#1{\mathcal{#1}}

\def\eql{=}

\def\RR{\mathbb{R}}

\def\rmi{{\rm i}}

\newcommand{\fg}{\mathfrak{g}}

\def\diag{{\rm diag}}

\def\SO{{\rm SO}}
\def\rU{{\rm U}}

\def\SU{{\rm SU}}

\def\rE{{\rm E}}

\def\sl{\frak{sl}}

\def\so{\frak{so}}
\def\su{\frak{su}}
\def\u{\frak{u}}

\def\fg{\frak{g}}
\def\fh{\frak{h}}

\def\bfs#1{{\boldsymbol{#1}}}

\def\cals#1{\mathcal{#1}}

\def\S#1#2#3#4#5#6#7{{\tt S#1#2#3#4#5#6#7}}

\numberwithin{equation}{section} 
\numberwithin{table}{section}
\numberwithin{figure}{section}

\def\rmi{i }

\begin{document}  
%%%%%%%%%%%%%%%%%%%%%%%%%%%%%%

\begin{titlepage}
 
\medskip
\begin{center} 
{\Large \bf A  New $\mathcal{N}=1$ AdS$_4$ Vacuum of Maximal Supergravity}

\bigskip
\bigskip
\bigskip
\bigskip

{\bf Nikolay Bobev,${}^{(1)}$ Thomas Fischbacher,${}^{(2)}$ and Krzysztof Pilch${}^{(3)}$   \\ }
\bigskip
${}^{(1)}$
Instituut voor Theoretische Fysica, KU Leuven,\\ 
Celestijnenlaan 200D, B-3001 Leuven, Belgium
\vskip 5mm
${}^{(2)}$ Google Research\\
Brandschenkestrasse 110, 8002 Z\"urich, Switzerland
\vskip 5mm
${}^{(3)}$ Department of Physics and Astronomy \\
University of Southern California \\
Los Angeles, CA 90089, USA  \\
\bigskip
\tt{nikolay.bobev@kuleuven.be, tfish@google.com, pilch@usc.edu}  \\
\end{center}

\bigskip
\bigskip

\begin{abstract}

\noindent The recent comprehensive numerical study  of  critical points of the scalar potential of four-dimensional  $\mathcal{N}=8$, $\SO(8)$ gauged supergravity using Machine Learning software  in \cite{Comsa:2019rcz} has led to a discovery of a new $\mathcal{N}=1$ vacuum with a triality-invariant $\SO(3)$ symmetry. Guided by the numerical data for that point, we obtain a consistent  $\SO(3)\times \ZZ_2$-invariant truncation of the $\cals N=8$ theory to an $\mathcal{N}=1$ supergravity with three chiral multiplets. Critical points of the truncated scalar potential include both the   $\mathcal{N}=1$ point as well as two new non-supersymmetric and perturbatively unstable points not found by previous searches.   Studying the structure of the submanifold of $\SO(3)\times \ZZ_2$-invariant supergravity scalars, we find that it has a simple interpretation as a submanifold of the 14-dimensional $\mathbb{Z}_2^3$-invariant scalar manifold $(\SU(1,1)/{\rm U}(1))^7$, for which we find a rather remarkable superpotential whose structure matches the single bit error correcting $(7,4)$ Hamming code. This 14-dimensional scalar manifold contains approximately one quarter of the known critical points. We also show that there exists a smooth supersymmetric domain wall which interpolates between the new $\mathcal{N}=1$ AdS$_4$ solution and the maximally supersymmetric AdS$_4$ vacuum. Using holography, this result indicates the existence of an $\mathcal{N}=1$ RG flow from the ABJM SCFT to a new strongly interacting conformal fixed point in the IR.
\end{abstract}

\vfill

\end{titlepage}

%%%%%%%%%%%%%%%%%%%%%%%%%%%%%%%%%%%%%
\newpage
\setcounter{tocdepth}{2}
\tableofcontents

%%%%%%%%%%%%%%%%%%%%%%%%%%%%%%%%%%%%%
\section{Introduction}
\label{sec:Introduction}
%%%%%%%%%%%%%%%%%%%%%%%%%%%%%%%%%%%%%

The four-dimensional  gauged supergravity of de Wit and Nicolai \cite{deWit:1982bul} has proven to be a remarkably rich theory  with a plethora of applications. It has an $\SO(8)$ gauge group and the maximal $\mathcal{N}=8$ supersymmetry. Much of the interesting physics in this theory arises from a highly nontrivial potential for the 70 scalar fields. It is perhaps fair to say that unlocking that physics is tantamount to understanding the structure of the potential. In particular, it has been a long standing problem to determine all of its critical points, which lead to  AdS$_4$ vacuum solutions of the theory. 

A systematic study of the nontrivial critical points, that is other than the maximally supersymmetric $\SO(8)$-invariant one with vanishing scalar fields, was initiated in \cite{Warner:1983du,Warner:1983vz}, where several AdS$_4$ vacua were found by imposing certain symmetry constraints and thus effectively reducing the 70-dimensional scalar manifold to a smaller one, which could be fully analyzed. 

As shown in \cite{deWit:1984nz,deWit:1986oxb,Nicolai:2011cy}, the four-dimensional $\SO(8)$ gauged supergravity\footnote{Throughout this paper, the SO(8) gauged supergravity means the original de Wit-Nicolai theory \cite{deWit:1982bul}.}  is   a consistent truncation of the eleven-dimensional supergravity on $S^7$. Therefore, classifying the critical points of the four-dimensional theory amounts to finding a large class of AdS$_4$ equilibria of eleven-dimensional supergravity with internal space that is topologically $S^7$.

Those AdS$_4$ vacua of the $\SO(8)$ gauged supergravity and their uplifts to eleven dimensions are also interesting from the point of view of holography. Indeed, such backgrounds are  dual to conformal field theories arising on the worldvolume of coincident M2-branes. The most studied and well-understood example is the ABJM theory \cite{Aharony:2008ug}, and its BLG version \cite{Bagger:2007jr,Gustavsson:2007vu}, which has maximal supersymmetry. Another example is the so called mABJM SCFT which has $\mathcal{N}=2$ supersymmetry and arises as a particular mass deformation of ABJM \cite{Benna:2008zy,Klebanov:2008vq,Jafferis:2011zi}.

The vacuum structure of the $\SO(8)$ gauged supergravity suggests that there are also conformal phases of M2-branes with $\mathcal{N}=1$ and $\mathcal{N}=0$ supersymmetry. In particular, the ${\rm G}_2$-invariant and the $\rU(1)\times\rU(1)$-invariant critical points found in \cite{Warner:1983vz} and  \cite{Fischbacher:2010ec}, respectively, preserve $\mathcal{N}=1$ supersymmetry. The $\SO(3)\times\SO(3)$-invariant critical point found in \cite{Warner:1983du} has no supersymmetry but it  is perturbatively stable \cite{Fischbacher:2010ec}. There is a conjecture that all non-supersymmetric AdS$_4$ vacua are unstable \cite{Ooguri:2016pdq}. However, it has not been explicitly shown how the non-perturbative instability of the $\SO(3)\times\SO(3)$-invariant critical point might arise.  

Due to the low amount of supersymmetry not much is known about these $\mathcal{N}=0$ and $\mathcal{N}=1$ strongly interacting CFTs. Nevertheless, it is interesting to understand whether there are any other supersymmetric or non-supersymmetric perturbatively stable critical points of the $\SO(8)$ gauged supergravity since this will amount to non-trivial predictions for the IR phases of the ABJM theory.

The consistent truncation of the maximal  supergravity using  a suitable symmetry as introduced by Warner \cite{Warner:1983du,Warner:1983vz} has remained  the   cornerstone for the analytic studies of the potential since 1983.  Between 2008 and 2010,  there has been also a considerable progress in developing numerical techniques to search for the critical points in the full 70-parameter space. Those methods were used by one of us to explore the vacuum structure of maximal gauged supergravity theories in three dimensions \cite{Fischbacher:2008zu,Fischbacher:2003rb} and then ported to four dimensions in \cite{Fischbacher:2009cj,Fischbacher:2010ki,Fischbacher:2011jx}. In particular,  a new $\cals N=1$ supersymmetric critical point $\tt S1200000$\footnote{Following  \cite{Fischbacher:2011jx}, we label the critical points by the first 7 digits of  the critical value of the potential.}   was discovered in \cite{Fischbacher:2009cj} and, using the numerical data as a guide, subsequently confirmed analytically in \cite{Fischbacher:2010ec}. 

A new method for determining critical points in gauged supergravities based on the embedding tensor formalism (see, e.g.,\ \cite{Trigiante:2016mnt}, and the references therein) was proposed in \cite{Dibitetto:2011gm} and \cite{DallAgata:2011aa,Inverso:2014kzl} in 2011. While this method has not yet led to any new stable vacua of the potential of the de~Wit-Nicolai theory,\footnote{See, however, the construction of a new SO(4)-invariant point in \cite{Borghese:2013dja}.} it has been used (see, e.g., \cite{DallAgata:2011aa,Borghese:2013dja,Borghese:2012qm,Borghese:2012zs}) to obtain new analytic results for critical points in the deformed SO(8) gauged supergravities constructed in \cite{DallAgata:2012mfj}. It also provided for a simple proof in \cite{Gallerati:2014xra} that these theories, including the original SO(8) gauged supergravity, have no supersymmetric vacua with $\cals N>2$ except for the maximally supersymmetric one.

Recently, a new numerical approach based on Machine Learning (ML) software libraries, such as Google's  TensorFlow~\cite{Abadi2016}, was employed in \cite{Comsa:2019rcz} to simplify the analysis of the potential resulting in the total of 192 critical points together with a precise information about those points  that includes  the mass spectra of small fluctuations  and unbroken (super)symmetries. It is expected that this list of   critical points should be  nearly complete.

Perhaps the most interesting result of the search in  \cite{Comsa:2019rcz}  is a discovery of yet another $\cals N=1$ supersymmetric critical point, $\tt S1384096$, which is invariant under a triality symmetric $\SO(3)$ subgroup of the $\SO(8)$ gauge group. Moreover, this point gives rise to the only new AdS$_4$ solution that is perturbatively stable. Therefore, it is most interesting to understand how to construct it using a more analytic approach. This is our goal in this paper.

The numerical data for the $\cals N=1$ critical point, $\tt S1384096$, in \cite{Comsa:2019rcz} point towards additional symmetry, which we identify as a discrete  $\ZZ_2$ subgroup of the $\SO(8)$ gauge group. The resulting $\SO(3)\times \ZZ_2$-invariant truncation of the $\cals N=8$ supergravity can be constructed analytically. Its bosonic sector   consists  of the metric and three complex scalar fields. Despite the small scalar sector, the potential in this truncation has 15 inequivalent critical points with  $\tt S1384096$  amongst them. Surprisingly,  two of those 15  critical points,
$\tt S2096313$ and $\tt S2443607$,  were not found by the numerical search in \cite{Comsa:2019rcz} and thus they represent new AdS$_4$ equilibria. However, they are not supersymmetric and are perturbatively unstable. 

We also study some of the properties of the supersymmetric point $\tt S1384096$ in more detail. In particular, we compute the mass spectrum of excitations for all bosonic and fermionic fields of the $\mathcal{N}=8$ supergravity around this point. Using holography, we map it to the spectrum of operators in the dual $\mathcal{N}=1$ three-dimensional SCFT, which are then organized   into multiplets of $\mathcal{N}=1$ superconformal symmetry. Our explicit analytic construction of the  truncation with three complex scalar fields also allows us  to initiate the study of the web of holographic RG flows connecting the four supersymmetric critical points in this model. We find explicit domain wall solutions which interpolate between $\tt S1384096$ and the maximally supersymmetric critical point of the $\mathcal{N}=8$ supergravity.

In the next section we discuss two $\SO(3)$- and $\SO(3)\times \mathbb{Z}_2$-invariant truncations of the maximal supergravity. In Section~\ref{sec:newpoint} we show that the new $\mathcal{N}=1$ AdS$_4$ solution found in \cite{Comsa:2019rcz} corresponds to a critical point in the $\SO(3)\times \mathbb{Z}_2$-invariant truncation. In Section~\ref{sec:otherpoints}, we apply the same numerical technique as in \cite{Comsa:2019rcz} to the potential in the $\SO(3)\times \mathbb{Z}_2$-invariant truncation and  find the total of 15 critical points that also include two  non-supersymmetric and perturbatively unstable ones that were missed by previous searches. We show how other well-known critical points arise in our truncation. In Section~\ref{sec:flows} we perform a preliminary study of the holographic RG flows to the new $\cals N=1$ point. In Section~\ref{sec:Z23} we present an $\mathcal{N}=1$ supergravity truncation of the maximal supergravity with 7 complex scalar fields which contains all perturbatively stable critical points. We conclude with some comments  in Section~\ref{sec:conclusions}. Some group theory  details,  the full spectrum of four-dimensional supergravity fields around the new  $\mathcal{N}=1$ AdS$_4$ vacuum,  as well as more details on the two new non-supersymmetric critical points are given in the appendices.

%%%%%%%%%%%%%%%%%%%%%%%%%%%%%%%%%%%%%
\section{A consistent truncation}
\label{sec:truncation}
%%%%%%%%%%%%%%%%%%%%%%%%%%%%%%%%%%%%%

The starting point of our analytic search for the new $\cals N=1$ critical point,  $\tt S1384096$, is the precise information about its symmetry that is given, together with the numerical data for the position of the point and the spectrum of supergravity fluctuations, in \cite{Comsa:2019rcz}. Specifically, we know that $\tt S1384096$ lies within an $\SO(3)$-invariant sector of the 70-dimensional scalar manifold, $\rm E_{7(7)}/(SU(8)/\mathbb{Z}_2)$,   of the $\cals N=8$ supergravity. The particular $\SO(3)$ symmetry is identified as the triality invariant  subgroup of the $\SO(8)$ gauge group specified by the following branchings of the three fundamental representations:
\begin{equation}\label{}
\bfs 8_{v,s,c}\quad \longrightarrow\quad \bfs 3 \oplus \bfs 3\oplus \bfs 1\oplus \bfs 1\,.
\end{equation}
Using the standard group theory summarized in Appendix~\ref{appendixA}, 
this completely determines the embedding of that $\rm SO(3)$ into both  $\rm SO(8)$ and $\rm E_{7(7)}$ at the level of Lie algebras:
\begin{equation}\label{embed}
\so(8)\supset \so(3)\times \frak u(1)\times \frak u(1)\quad  \text{and}\qquad \frak e_{7(7)}\supset \so(3)\times \fg_{2(2)}\times \su(1,1)\,.
\end{equation}
Indeed, those embeddings are confirmed by the $\rU(1)\times \rU(1)$ unbroken gauge symmetry and the 
the presence  of $8+2=10$ scalar  fluctuations\footnote{See, Table~\ref{tbl-m0} in Appendix~\ref{AppendixB}.} that are $\SO(3)$ singlets in the $\cals N=8$ supergravity spectrum around that point.  

It follows from \eqref{embed} that 
 the scalar manifold spanned by the $\rm SO(3)$-invariant scalars is the coset\footnote{Incidentally, the first factor in the coset has appeared in the $\SU(3)$-invariant truncation of the type IIB supergravity \cite{Ferrara:1989ik,Bodner:1989cg} and in the $\SO(3)$-invariant consistent truncation of the maximal five-dimensional $\SO(6)$ supergravity  \cite{Pilch:2000fu}.}
\begin{equation}\label{eq:G2coset}
\ds\frac{{\rm G}_{2(2)}}{\SO(4)} \times \ds\frac{\SU(1,1)}{\rU(1)}\,.
\end{equation}
In fact, keeping track of all invariant fields in the $\cals N=8$ supergravity, one finds that the (consistent) $\SO(3)$-invariant truncation is to a four-dimensional $\mathcal{N}=2$ gauged supergravity coupled to an Abelian vector multiplet and two hypermultiplets. The scalars in the hyper and the vector multiplets parametrize the first and second factor in \eqref{eq:G2coset}, respectively. While considerably simpler than the full $\cals N=8$ theory, this truncation is still too complicated to effectively work with.

The crucial hint that allows us to further simplify the analytic search for  $\tt S1384096$ comes from  the numerical values of the scalar fields at that point. It has been  observed in  \cite{Comsa:2019rcz} that one can specify the position of this point in terms of 6 independent scalars when using a certain parametrization of the $\rE_{7(7)}/\SU(8)$ coset. This suggests the presence of an additional discrete symmetry that might allow  further truncation of the scalar manifold to a smaller subspace. 

To identify that discrete symmetry we will use the standard parametrization of the scalar manifold of the $\cals N=8$ theory in which the scalar 56-bein is given by \cite{Cremmer:1979up,deWit:1982bul}
\begin{equation}\label{56bein}
\cals V\equiv \left(\begin{matrix}
u_{ij}{}^{IJ} & v_{ijIJ} \\ v^{klIJ} & u^{kl}{}_{KL}
\end{matrix}\right)\eql  \exp \left(\begin{matrix}
0 & -\coeff 1 4 \sqrt 2\,\phi_{ijkl}\\
 -\coeff 1 4 \sqrt 2\,\bar\phi^{ijkl} & 0
\end{matrix}\right)\in {\rm E_{7(7)}}\,,
\end{equation}
where the scalar fields, $\phi_{ijkl}$, are components of a  complex selfdual 4-form in $\RR^8$,
\begin{equation}\label{phiforms}
\Phi\eql {1\over 24}\sqrt{2} \,\phi_{ijkl}\,dx^i\wedge dx^j\wedge dx^k\wedge dx^l\,,
\end{equation}
and the indices $i,j,\ldots$ transform  in the $\bfs 8_v$ representation of $\SO(8)$. To simplify the notation, the wedge product above will be denoted by $dx^{ijkl}$. 

We take the $\rm SO(3)$ symmetry of the truncation to act diagonally  on the indices $(123)$ and $(456)$ with the singlets $(7)$ and $(8)$. The invariant forms under this action are spanned by:
\begin{equation}\label{critsub}
\begin{split}
\Phi^{(1)} & \eql \,\omega_1\,\left( dx^{1267}-dx^{1357}+dx^{2347}\right)
+\overline\omega_1\,\left(dx^{1568}-dx^{2468}+dx^{3458}\right)\,,\\
\Phi^{(2)} & \eql \omega_2\,dx^{4567}+\overline \omega_2\,dx^{1238}\,,\\
\Phi^{(3)} & \eql \omega_3\,\left(dx^{1245}+dx^{1346}+dx^{2356}\right)+
\overline\omega_3\,\left(dx^{1478}+dx^{2578}+dx^{3678}\right)\,,
\end{split}
\end{equation}
and
\begin{equation}\label{othersub}
\begin{split}
\Phi^{(4)}& \eql \vartheta_1\,\left(dx^{1268}-dx^{1358}+dx^{2348}\right)-
\overline\vartheta_1\,\left(dx^{1567}-dx^{2467}+dx^{3457}\right)\,,\\
\Phi^{(5)}& \eql \vartheta_2\,dx^{4568}-\overline\vartheta_2\,dx^{1237}\,,
\end{split}
\end{equation}
where
\begin{equation}\label{defss}
\omega_a\eql s_{2a-1}+\rmi\, s_{2a}\,,\quad a=1,2,3\,,\qquad \vartheta_j\eql t_{2j-1}+\rmi\, t_{2j}\,,\quad j=1,2\,.
\end{equation}
Each $\Phi^{(\alpha)}$ contains a scalar and a pseudoscalar that parametrize an $\rm SU(1,1)/U(1)$ coset. The $\rm SU(1,1)$ subalgebras of $\rm E_{7(7)}$ generated by $\Phi^{(1)}$, $\Phi^{(2)}$ and $\Phi^{(3)}$ mutually commute as do  the two subalgebras corresponding to $\Phi^{(4)}$ and $\Phi^{(5)}$.
The  $\Phi^{(1)}$,   $\Phi^{(2)}$, $\Phi^{(4)}$ and  $\Phi^{(5)}$ correspond to the noncompact generators of $\rm G_{2(2)}$ and $\Phi^{(3)}$ to the  $\rm SU(1,1)$ that commutes with that $\rm G_{2(2)}$.

This truncation contains the smaller $\SU(3)$-invariant truncation used by Warner \cite{Warner:1983vz} and more recently discussed in \cite{Bobev:2010ib}. It is obtained by setting $\omega_1=-\omega_2$ and $\vartheta_1=-\vartheta_2$, which results in the noncompact group $\SU(2,1)\times \SU(1,1)$.

Using the numerical data for  $\tt S1384096$ in  \cite{Comsa:2019rcz}, we find that, modulo a suitable $\SO(8)$ gauge rotation, the critical point of the potential lies within  the 6-dimensional subspace spanned by $\Phi^{(1)}$, $\Phi^{(2)}$ and $\Phi^{(3)}$. We also find that the fluctuation of the gravitino field, $\psi_\mu{}^8$, remains massless as required by the unbroken $\cals N=1$ supersymmetry. 

Consider the following discrete symmetry
\begin{equation}\label{gS}
g_S:\quad (x^1,x^2,x^3,x^4,x^5,x^6,x^7,x^8)\quad \longrightarrow\quad (x^1,x^2,x^3,-x^4,-x^5,-x^6,-x^7,x^8)\,.
\end{equation}
Clearly,  $g_S$ is an $\rm SO(8)$ rotation that does not belong to the $\rm SO(3)$ symmetry subgroup. Under the action of $g_S$, the forms in \eqref{critsub} are even while the ones in \eqref{othersub} are odd, and the correct supersymmetry is preserved. This provides us with an additional discrete symmetry for the truncation to 6 scalar fields, where the truncation simply  amounts to setting
\begin{equation}\label{ttrunc}
t_1\eql t_2\eql t_3\eql t_4=0\,.
\end{equation}
 In the next section we will confirm directly that the critical point  $\tt S1384096$ indeed lies in the ${\rm SO(3)}\times \ZZ_2$ invariant sector, where  the $\ZZ_2$ is generated by the $\rm SO(8)$ rotation $g_S$.

To summarize, we have been led by the group theory and  numerical data to consider a $\SO(3)\times \ZZ_2$-invariant truncation of the $\cals N=8$ $d=4$ supergravity with the scalar coset 
\begin{equation}\label{eq:su11coset}
 \ds\frac{\SU(1,1)}{\rU(1)}\times \ds\frac{\SU(1,1)}{\rU(1)} \times \ds\frac{\SU(1,1)}{\rU(1)}\,.
\end{equation}
The first two factors above are embedded in the first factor in \eqref{eq:G2coset} and the last factor above corresponds to  the second factor in \eqref{eq:G2coset}. The resulting theory is an $\cals N=1$ $d=4$ supergravity coupled to  3 scalar multiplets.  The  $\cals N=8$ fields that remain in this truncation are indicated by the star in Tables~\ref{tbl-m0}-\ref{tbl-m32}.

To recast the bosonic sector of this $\cals N=1$ $d=4$ supergravity in a canonical form \cite{Freedman:2012zz}, 
we use the usual complex coordinates, $z_a$, on each   $\rm SU(1,1)/U(1)$. First set
\begin{equation}\label{}
s_{2a-1}+\rmi s_{2a}\eql \lambda_a e^{\rmi\,\varphi_a}\,,\qquad  a=1,2,3\,,
\end{equation}
with the $s_a$ given in \eqref{defss}, and then define
\begin{equation}\label{}
z_a\eql \tanh(\hbox{$1\over 2$}\lambda_a)\,e^{\rmi\,\varphi_a}\,,\qquad a=1,2,3\,.
\end{equation}

In this parametrization of the scalar fields, the bosonic  Lagrangian of the truncated four-dimensional supergravity is given by
\begin{equation}\label{eq:Lag4d}
\mathcal{L} ={ 1\over 2}  R - \mathcal{K}^{a\bar{b}}\partial_{\mu}z_a\partial^{\mu}\bar{z}_{\bar{b}} -g^2 \mathcal{P}\,.
\end{equation}
The scalar kinetic term  is determined by the K\"ahler metric of the coset \eqref{eq:su11coset} with K\"ahler potential
\begin{equation}\label{eq:cKdef}
\mathcal{K} = -3\log(1-z_1\bar{z}_1)-\log(1-z_2\bar{z}_2)-3\log(1-z_3\bar{z}_3)\,,
\end{equation}
where the integer coefficients of the logarithms  are the embedding indices of the numerator $\rm SU(1,1)$'s in \eqref{eq:su11coset} in $E_{7(7)}$.
The K\"ahler metric and its inverse are defined by
\begin{equation}\label{eq:defKstuff}
\mathcal{K}_{a\bar{b}} = \partial_{a}\partial_{\bar{b}}\mathcal{K}\,, \qquad \mathcal{K}^{a\bar{b}} = (\mathcal{K}_{a\bar{b}})^{-1}\,.
\end{equation}
The potential can be succinctly written as 
\begin{equation}\label{eq:cPdef}
\mathcal{P} = 2e^{\mathcal{K}}\big(\mathcal{K}^{ a\bar{b}}\nabla_{a}\mathcal{W}\nabla_{\bar{b}}\overline{\mathcal{W}}-3\mathcal{W}\overline{\mathcal{W}}\,\big)\,,
\end{equation}
where the holomorphic superpotential is\footnote{The holomorphic  superpotential can be read-off from the component of the $A^1_{ij}$-tensor of the $\cals N=8$ supergravity  along the  unbroken supersymmetry.}
\begin{equation}\label{eq:cWdef}
\mathcal{W} = (z_3-1)(z_1^3z_2z_3^2+z_1^3z_2z_3+z_1^3z_2+3z_3z_1^2-3z_1z_2z_3-z_3^2-z_3-1)\,,
\end{equation}
and the K\"ahler covariant derivative is given by
\begin{equation}\label{}
\qquad \nabla_{a}(\cdot) = \partial_{a}(\cdot)+(\cdot)\partial_{a}\mathcal{K}\,.
\end{equation}

%%%%%%%%%%%%%%%%%%%%%%%%%%%%%%%%%%%%%
\section{The new $\cals N=1$ critical point}
\label{sec:newpoint}
%%%%%%%%%%%%%%%%%%%%%%%%%%%%%%%%%%%%%

In this section we will look first for those critical points of the potential  \eqref{eq:cPdef} that preserve the $\cals N=1$ (or more) supersymmetry of our  truncation. The expectation is that those points should include the  new supersymmetric  point, $\tt S1384096$. 

Supersymmetric critical points correspond to the ``covariant extrema'' of the holographic superpotential \eqref{eq:cWdef} satisfying
\begin{equation}\label{eq:nablaWeq}
\nabla_{a}\mathcal{W}=0\,,\qquad \nabla_{\bar{b}}\overline{\mathcal{W}}=0\,,\qquad a,\bar{b}=1,2,3\,.
\end{equation}
It is easy to check that any solution  to  \eqref{eq:nablaWeq} is also a critical point of the  potential  \eqref{eq:cPdef}.

In the domain $|z_a|<1$, \eqref{eq:nablaWeq}  unpack to the following system of septic polynomial equations
\begin{equation}\label{zsusyeqs}
\begin{split}
z_1^2 z_2+2 z_1 z_3-z_2 z_3+z_1^2 z_2 z_3+z_1^2 z_2 z_3^2-\bar z_1(1+z_3 -z_1^2 z_3+2 z_1 z_2 z_3+z_3^2) & \eql 0\,,
\\
z_1^3-3 z_1 z_3+3 z_1 z_3^2-z_1^3 z_3^3-\bar z_2(1 -3 z_1^2 z_3+3 z_1^2 z_3^2-z_3^3)& \eql 0\,,
\\
z_1^2 -z_1 z_2 -2 z_1^2 z_3+2 z_1 z_2 z_3+z_3^2 -z_1^3 z_2 z_3^2\hspace{2 in} &\\ -\bar z_3(
1-z_1^3 z_2-2 z_1^2 z_3+2 z_1 z_2 z_3+z_1^2 z_3^2-z_1 z_2 z_3^2)& \eql 0\,,
\end{split}
\end{equation}
plus  the three complex conjugate equations.

Finding all solutions to the system \eqref{zsusyeqs} analytically, if feasible at all, will require techniques that go beyond what is employed in the present article. However, one can fully analyze \eqref{zsusyeqs} using standard numerical routines such as $\tt NSolve[\cdot]$ in Mathematica \cite{Mathematica}. In this way we recover three known supersymmetric critical points with the  $\SO(8)$, $\rm G_2$ and $\SU(3)\times \rU(1)$ invariance, respectively. As we discuss in Section~\ref{ssec:subtr}, each can be found analytically by performing a further truncation  that reduces \eqref{zsusyeqs} to a simpler system. 

We also find four numerical solutions at approximately
\begin{align}\label{num4pts}
z_1 & \eql 0.1696360 \pm 0.1415740 \,\rmi\,, &  z_1 & \eql -0.1696360 \mp 0.1415740 \,\rmi\,,\notag \\
z_2 & \eql 0.4833214 \pm 0.3864058  \,\rmi\,,  & z_2 & \eql -0.4833214 \mp 0.3864058  \,\rmi\,, \\
z_3 & \eql -0.3162021 \pm 0.5162839\,  \rmi\,, & z_3 & \eql -0.3162021 \pm 0.5162839\,  \rmi\,.
\notag
\end{align}
The value of the potential at these points is
\begin{equation}\label{eq:potnumSO3}
\mathcal{P} \approx -13.840964\,,
\end{equation}
which is the same as at  $\tt S1384096$  in  \cite{Comsa:2019rcz}.  

As we discuss in more detail in Appendix~\ref{AppendixC}, there is a residual action of the $\SO(8)$ gauge symmetry on the coset \eqref{eq:su11coset}. Indeed, solutions in the two columns in \eqref{num4pts} are related by the rotation, $g_H$, defined in \eqref{dtrans}. In turn, solutions within each column are related by complex conjugation. This degeneracy is expected given the corresponding invariance of the system of equations in  \eqref{eq:nablaWeq} or, equivalently, \eqref{zsusyeqs}. However, those complex conjugate solutions are {\it not} related by an  $\SO(8)$ rotation and thus represent two distinct critical points of the potential in the $\cals N=8$ supergravity.\footnote{Since such ``conjugate'' critical points have the same values of the potential and the same mass spectra of fluctuations around them, they were identified as a single point in the numerical searches \cite{{Fischbacher:2011jx},Comsa:2019rcz}.} To complete the identification of the solutions \eqref{num4pts} with the new supersymmetric point,  $\tt S1384096$, and its complex conjugate, $\tt \overline S1384096$, in \cite{Comsa:2019rcz}, one can also perform an explicit change of variables \eqref{changvar} accompanied by an $\SO(8)$ rotation.

The  critical values \eqref{num4pts} of the coordinates, $z_a$, $a=1,2,3$, can be efficiently determined to an arbitrary precision from the roots of the following system of integer coefficient polynomials:\footnote{See, Section~\ref{numerics}. In Appendix~\ref{App:filespol}, we also give the minimal polynomials for the complex coordinates \eqref{polz1}-\eqref{polz3}.} 
\begin{equation}\label{polz1}
\begin{split}
%P_{|z_1|}(x) & \eql x^{24}+4\, x^{22}-302\, x^{20}-3404\, x^{18}-11985\, x^{16}-38328 \,x^{14}-51972 \,x^{12}\\ 
% &\quad -38328\, x^{10} -11985\, x^8-3404\, x^6-302 \,x^4+4 x^2+1 \,,\\[6 pt]
P_{\Re\, z_1}(x) & \eql 256\, x^{24}+5632 \,x^{22}+78592 \,x^{20}-2135808 \,x^{18}-543360 \,x^{16}-4684032 \,x^{14}\\
&\quad -12045600 \,x^{12}-15419808 \,x^{10}-6033744
   \,x^8-1553904 \,x^6-222264 \,x^4\\
   &\quad -17496 \,x^2+729\,,\\ 
P_{\Im\,z_1}(y) & \eql 16  y^{12}-96  y^{11}+144  y^{10}+848  y^9-944  y^8+2000  y^7-3504  y^6+3456  y^5\\
&\quad -2028  y^4+740  y^3-164  y^2+20  y-1 \,.  
   \end{split}
\end{equation}
\begin{equation}\label{polz2}
\begin{split}
%P_{|z_2|}(x) & \eql  487\, x^{24}+1072 \,x^{22}+1846 \,x^{20}+208\, x^{18}-30327\, x^{16}-56832\, x^{14}+7092\, x^{12}\\ 
%&\quad -56832\, x^{10}-30327\, x^8+208 \,x^6+1846 \,x^4+1072
%   x^2+487 \,,\\[6 pt]
P_{\Re\, z_2}(x) & \eql 60715264 \,x^{24}-256862720 \,x^{22}+937708288 x^{20}-2138845440 \,x^{18}\\
&\quad +3067400064 \,x^{16}-2992061952 \,x^{14}+1409137632
   \,x^{12}-388837152 \,x^{10}\\ 
&\quad +229269744 \,x^8-95418000 \,x^6+4147848 \,x^4+1627128 \,x^2+59049\,,\\ 
P_{\Im\,z_2}(y)&\eql
7792  y^{12}+4320  y^{11}+26256  y^{10}+16832  y^9+62032  y^8+107504  y^7+70872  y^6\\ &\quad +37872  y^5+14172  y^4+19880  y^3+4900
    y^2-1372  y-2401 \,,
   \end{split}
\end{equation}
\begin{equation}\label{polz3}
\begin{split}
%P_{|z_3|}(x) & \eql 3 \,x^{24}-104 \,x^{22}+838 \,x^{20}+1368 \,x^{18}-1395 \,x^{16}-14448 \,x^{14}-4524 \,x^{12}\\
%&\quad -14448 \,x^{10}-1395 \,x^8+1368 \,x^6+838 \,x^4-104 \,x^2+3\,,\\[6 pt]
P_{\Re\, z_3}(x) & \eql 16 \,x^{17}+96 \,x^{16}+496 \,x^{15}+672 \,x^{14}+456 \,x^{13}-1584 \,x^{12}-1384 \,x^{11}-816 x^{10}\\
&\quad +1388 \,x^9-1512 \,x^8+462 \,x^7+2028 \,x^6+537
   \,x^5+810 \,x^4-819 \,x^3-639 \,x^2\\ &\quad -180 \,x-27\,,\\ 
P_{\Im \,z_3}(y) & \eql  768  y^{24}-48128  y^{22}+1018112  y^{20}-2517248  y^{18}+4496192  y^{16}-8476736  y^{14}\\ &\quad +7496864  y^{12}-8223008
    y^{10}+4957568  y^8-1487120  y^6+233460  y^4\\ &\quad -18900  y^2+675\,.  
\end{split}
\end{equation}
Polynomials $P_{\Re\,z_a}(x)$ and $P_{\Im\,z_a}(y)$, $a=1,2$, have precisely two real roots, $\pm x_a^*$, and one real root, $y_a^*$, respectively, such that $z_a=x_a^*+i\,y_a^*$, $x_a^*,y_a^*>0$,  lie in the unit disk. Similarly, $P_{\Re\,z_3}(x)$ and $P_{\Im\,z_3}(y)$ have one real root, $x_3^*$, and two real roots, $\pm y_3^*$, $y_3^*>0$, with $|z_3^*|<1$. Then 
\begin{equation}\label{sol4W}
(z_1^*,z_2^*,z_3^*)\,,\qquad (-z_1^*,-z_2^*,z_3^*)\,,\qquad (\bar z_1^*,\bar z_2^*,\bar z_3^*)\,,\qquad  (-\bar z_1^*,-\bar z_2^*,\bar z_3^*)\,,
\end{equation}
are the four solutions to \eqref{eq:nablaWeq} that we found above in \eqref{num4pts}. By acting on these solutions with the $\rm SO(8)$ rotation $g_C$ defined in \eqref{dtrans}, we obtain 4 additional solutions that exhaust  the critical points of the potential \eqref{eq:cPdef} with this critical value. Those 8 critical points of \eqref{eq:cPdef} represent two different critical points of the $\cals N=8$ supergravity.

The exact value of the potential at these critical points is given by the negative real root of the following polynomial \cite{Comsa:2019rcz},
\begin{equation}\label{}
5^{15}\,v^{12}-(2^8\cdot 3^4\cdot 7\cdot 53\cdot 107\cdot 887\cdot
   1567)\,v^8+(2^{15}\cdot 3^{17}\cdot 210719)\,v^4- 2^{20}\cdot 3^{30}\eql 0\,.
\end{equation}

The mass spectra of the fermion and scalar fluctuations of $\cals N=8$ supergravity are the same around each of the points and have been obtained as part of the numerical search in \cite{Comsa:2019rcz}. Those results are summarized in  Tables~ \ref{tbl-m12}, \ref{tbl-m32} and \ref{tbl-m0} in Appendix~\ref{AppendixB}. In particular, the presence of a single  unit mass gravitino mode in the spectrum  shows that there is $\cals N=1$ unbroken supersymmetry. In  Appendix~\ref{AppendixB} we also find the masses of the fluctuations of the vector fields, which allows us to verify explicitly that the entire spectrum of  operators in the dual three-dimensional superconformal  field theory can be arranged into multiplets of the $\cals N=1$ superconformal algebra, $\frak o\frak s\frak p(1|4)$. This provides a nontrivial consistency test  for the calculation of the spectra and of the unbroken supersymmetry.

%%%%%%%%%%%%%%%%%%%%%%%%%%%%%%%%%%%%%
\section{All critical points of the truncated potential}
\label{sec:otherpoints}
%%%%%%%%%%%%%%%%%%%%%%%%%%%%%%%%%%%%%

Although the truncated potential \eqref{eq:cPdef} can be written in a closed analytic form, to determine all of its critical points we have to resort to numerical methods outlined in Section~\ref{numerics}. Here, we   summarize the results of that search, which yielded critical points with 15 different values of the cosmological constant, including the two new ones that were not captured by the search in  \cite{Comsa:2019rcz}.

Given a critical point  at  $(z_1,z_2,z_3)$, the reality of the potential implies that there is another (conjugate)  critical point at $(\bar z_1,\bar z_2,\bar z_3)$. In Appendix~\ref{AppendixC}, we argue that unless $z_3$ is real, the two points are not related by an $\SO(8)$ rotation and thus represent two distinct critical points of the potential in the $\cals N=8$ supergravity. In addition, we show that each point in the coset  \eqref{eq:su11coset} lies on an orbit of the discrete subgroup of $\SO(8)$ that preserves the coset. This discrete subgroup is generated by the rotations  $g_H$ and $g_C$ defined in \eqref{dtrans}. Generically, the corresponding orbit through a point  $(z_1,z_2,z_3)$ consists of 4 points:
\begin{equation}\label{dorbit}
(z_1,z_2,z_3)\,,\qquad  (-z_1,-z_2,z_3)\,, \qquad (\bar z_1,\bar z_2,z_3)\,,\qquad (-\bar z_1,-\bar z_2,z_3)\,,
\end{equation}
obtained by applying the rotations $1$, $g_H$, $g_C$ and $g_Hg_C$, respectively. Clearly, when $z_1$ and $z_2$ ($z_1z_2\not=0$) are both either real or imaginary, that orbit  degenerates to just two points. For some points we also find additional discrete $\rm SO(8)$ rotations \eqref{comrot} that preserve the coset \eqref{eq:su11coset} at that particular point giving rise to additional critical points of the potential \eqref{eq:cPdef}.

The end result is that for each critical value of the potential \eqref{eq:cPdef} at $(z_1,z_2,z_3)$, there are either two orbits or a single orbit of critical points, namely \eqref{dorbit} and its complex conjugate, or just \eqref{dorbit} when $z_3$ is real.  In $\cals N=8$ supergravity those $\SO(8)$ orbits correspond two ``conjugate''  critical points, $\tt  Sn_1\ldots n_7$ and $\tt\overline Sn_1\ldots n_7$, or a single point, $\tt Sn_1\ldots n_7$, respectively.

All points in this section have at least an $\SO(3)$ symmetry, which is the continuous symmetry of  the truncation. We should note that there are other critical points in  \cite{Comsa:2019rcz} that are $\SO(3)$-invariant: 
$\S0847213$, $\S1075828$, $\S1195898$, $\S1271622$, $\S2503105$\,.
However, their symmetry is incompatible with the symmetry of our truncation since it corresponds to a different embedding of $\SO(3)$ in $\SO(8)$.

In the following we list all the critical points of the potential \eqref{eq:cPdef}. For each point we  give its location, the value  of the potential, the continuous symmetry, and the $\SO(8)$ rotations for the orbit(s). All but two of those points were discovered in previous searches as indicated by the references where they first appeared. For all but two points the location  and the critical value of the potential are known in either a closed analaytic form or via a minimal polynomial. In some cases where the explicit analytic form is too involved, we do not list it in the text.

The mass spectra of scalar fluctuations around the known points can be found in \cite{Comsa:2019rcz}.\footnote{Also, see earlier work referred to in \cite{Comsa:2019rcz}.} For the two new points, the mass spectra are given in Appendix~\ref{appNew}. Perhaps unsurprisingly, the only perturbatively stable points, that is with the scalar masses satisfying the Breitenlohner-Freedman bound \cite{Breitenlohner:1982jf}, are the supersymmetric ones and the non-supersymmetric $\SO(3)\times \SO(3)$-invariant point, $\tt S1400000$. In particular, both new points are perturbatively unstable. Overall, for the solutions $\tt S0668740$, $\tt S0698771$, $\tt S0800000$, and $\tt S1424025$, all instabilities in this truncation are due to modes that are not $\SO(3)$ singlets and thus can be seen only within the full $\cals N=8$ supergravity. For the $\SU(4)$-invariant solution $\tt S0800000$ this observed previously in \cite{Bobev:2010ib}.

\subsection{The critical points}
\label{subsec:critpot}

\begin{itemize}[leftmargin=*]
\item [] $\bf S0600000$ \cite{deWit:1982bul}
\begin{equation}\label{so8point}
z_1\eql z_2\eql z_3\eql 0\,,
\end{equation}
\begin{equation}\label{}
\qquad \cals P\eql -6\,.
\end{equation}
\qquad 
Symmetry: $\SO(8)$,\quad  $\cals N=8$.\qquad Orbit: $\langle 1\rangle$.
\\[-6 pt]
%%%%%%%%%%%%%%%%

\item[] $\bf S0668740$ \cite{deWit:1984va}
\begin{equation}\label{eq:SO7pcrit}
z_1\eql -z_2\eql -z_3\eql {1\over 2}\left (3 + \sqrt 5 - \sqrt{10 + 6 \sqrt 5}\,\right)\approx 0.1985088\,, 
\end{equation}
\begin{equation}\label{}
\mathcal{P} = -2\cdot 5^{3/4} \approx -6.687403\,.
\end{equation}

\qquad 
Symmetry: $\rm SO(7)^+\,,\quad \cals N=0$.\qquad Orbit: $\langle 1,g_H\rangle$\,.
\begin{equation}\label{}
\cals P_{\Re\,z_{1,2,3}}(x)\eql x^4+6 x^3+6 x^2+6 x+1\,.
\end{equation}
%%%%%%%%%%%%%%%%

\item[]
$\bf S0698771~\&~\overline  S0698771$ \cite{Biran:1982eg,deWit:1983gs}
\begin{equation}\label{eq:SO7mcrit}
\begin{split}
z_1\eql - z_2\eql -z_3\eql -\rmi(2-\sqrt{5})\approx 0.2360680\rmi\,,
\end{split}
\end{equation}
\begin{equation}\label{}
\qquad 
\mathcal{P} = -\frac{5^{5/2}}{8} \approx -6.987712\,.
\end{equation}
\qquad
Symmetry: $\SO(7)^-\,,\quad \cals N=0$.\qquad Orbits: $\langle 1,g_H\rangle$.
\begin{equation}\label{}
P_{\Im\,z_{1,2,3}}(x)\eql x^2+4 x-1\,.
\end{equation}

%%%%%%%%%%%%%%%%%%%

\item[]
$\bf S0719157~\&~\overline S0719157$ \cite{deWit:1984nz}
\begin{equation}\label{eq:G2crit}
\begin{split}
z_1=- z_2=-z_3& =\frac{1}{4}\left(3+\sqrt{3}-3^{1/4}\sqrt{10}\right)\left(1-\rmi\, 3^{-1/4}\sqrt{2+\sqrt{3}}\right)\\ & \approx
0.1425648 + 0.2092695 \,i\,,\\ 
\end{split}
\end{equation}
\begin{equation}\label{}
\mathcal{P} = -\frac{2^{7/2}\cdot 3^{13/4}}{5^{5/2}} \approx -7.191576\,.
\end{equation}
\qquad 
Symmetry: $\rm G_2$,\quad $\cals N=1\,.$\qquad Orbits: $\langle 1,g_H,g_C,g_Hg_C\rangle$\,.
\begin{equation}\label{}
\begin{split}
P_{\Re \,z_{1,2,3}}(x) & \eql 8 x^4+24 x^3+24 x^2+24 x+3\,,\\
P_{\Im \,z_{1,2,3}}(x) & \eql64 x^8-704 x^6+240 x^4-32 x^2+1\,.
\end{split}
\end{equation}

%%%%%%%%%%%%%%%%%%

\item[]
$\bf S0779422$ \cite{Warner:1983vz}
\begin{equation}
\label{su3u1pt}
\begin{split}
z_1\eql -z_2\eql \rmi\,\sqrt{5-2\sqrt{6}} \approx 0.3178372\,i\,, \qquad  z_3=\sqrt{3}-2\approx -0.2679492 \,,
\end{split}
\end{equation}
\begin{equation}\label{}
\mathcal{P} = -\frac{9\sqrt{3}}{2} \approx -7.794229\,.
\end{equation}
\qquad
Symmetry: $\rm SU(3)\times U(1)$\,,\quad $\cals N=2$\,.\qquad Orbit: $\langle 1,g_H \rangle$\,.
\begin{equation}\label{}
\begin{split}
P_{\Im\,z_{1,2}}(x) & \eql x^4-10 x^2+1\,,\qquad P_{\Re\,z_3}(x)\eql x^2+4 x+1\,.
\end{split}
\end{equation}

%%%%%%%%%%%%%%%%%%%

\item[]
$\bf S0800000$  \cite{Warner:1983vz}
\begin{equation}\label{eq:SU4crit}
\begin{split}
z_1&=-z_2 \eql \rmi\,(\sqrt 2-1)\approx 0.4142136\,i \,, \quad  z_3=\bar{z}_3=0\,,\end{split}
\end{equation}
\begin{equation}\label{}
\qquad 
\mathcal{P} = -8\,.
\end{equation}
\qquad
Symmetry: $\rm SU(4)\,,\quad \cals N\eql 0$\,.\qquad Orbit: $\langle 1,g_H \rangle$\,.
\begin{equation}\label{}
P_{\Im\,z_{1,2}}(x)\eql x^2+2 x-1\,.
\end{equation}

%%%%%%%%%%%%%%%%%%

\item[]
$\bf S0869597$ \cite{Fischbacher:2011jx}
\begin{equation}\label{}
\begin{split}
z_1 & \eql i\,\sqrt{9+2 \sqrt{21}-2 \sqrt{41+9 \sqrt{21}}}\approx 0.1659702\rmi\,,\\[6 pt]
z_2 & \eql \frac{i}{67} \sqrt{7521+738 \sqrt{21}-2 \sqrt{11962961+2775249 \sqrt{21}}}\approx 0.4641278\rmi\,,\\[6 pt]
z_3 & \eql \frac{1}{4} \Big(-1-\sqrt{21}+\sqrt{2 \left(3+\sqrt{21}\right)}\Big)\approx -0.4220824\,,
\end{split}
\end{equation}
\begin{equation}\label{}
\cals P\eql -\frac{4}{5} \sqrt{54+14 \sqrt{21}}\approx  -8.695969\,.
\end{equation}
\qquad Symmetry: $\SO(3)\times \rU(1)\,,\quad \cals N=0$\,.\qquad Orbit $\langle 1,g_H \rangle$\,.
\begin{equation}\label{}
\begin{split}
P_{\Im\,z_1}(x) & \eql  x^8-36 x^6-10 x^4-36 x^2+1\,,\\
P_{\Im\,z_2}(x) & \eql 4489 x^8-30084 x^6+49190 x^4-30084 x^2+4489\,,\\
P_{\Re\,z_3}(x) & \eql x^4+x^3-3 x^2+x+1\,.
\end{split}
\end{equation}

%%%%%%%%%%%%%%%%%%%
\noindent
$\bf S0880733$ \cite{Fischbacher:2010ec}
\begin{equation}\label{}
\begin{split}
z_1& \eql z_2 \eql i\, \sqrt{3+2 \sqrt{3}-2 \sqrt{5+3 \sqrt{3}}}\approx 0.2789600\rmi\,,\\[6 pt]
z_3& \eql -\frac{1}{2}+\frac{3^{1/4}}{\sqrt{2}}-\frac{\sqrt{3}}{2}\approx -0.435421\,,
\end{split}
\end{equation}

\begin{equation}\label{}
\cals P\eql -2 \sqrt{9+6 \sqrt{3}} \approx -8.807339\,.
\end{equation}
\qquad
Symmetry: $\SO(4)\,,\quad \cals N=0$\,.\qquad Orbit: $\langle 1,g_H,g_R,g_Hg_R\rangle\,. $
\begin{equation}\label{}
\begin{split}
P_{\Im \,z_{1,2}}(x) & \eql x^8-12 x^6-10 x^4-12 x^2+1\,,\\
P_{\Re\,z_{3}}(x) & \eql x^4+2 x^3+2 x+1\,.
\end{split}
\end{equation}

%%%%%%%%%%%%%%%%%%
\item[]
$\bf S0983994~\&~\overline S0983994$ \cite{Fischbacher:2011jx}
\begin{equation}\label{}
z_1\approx 0.184246 \,,\qquad z_2\approx 0.5073269\,,\qquad z_3\approx 0.1331835+0.4676097\,i\,,
\end{equation}
\begin{equation}\label{}
\cals P\eql -5\cdot 15^{1/4}\approx -9.839948\,.
\end{equation}
\qquad
Symmetry: $\rm SO(3)\times \rU(1)\,,\qquad \cals N=0\,.$\qquad Orbits: $\langle 1,g_H\rangle$\,.
\begin{equation}\label{}
\begin{split}
P_{\Re\,z_1}(x) & \eql x^8-28 x^6-42 x^4-28 x^2+1\,,\\
P_{\Re\,z_2}(x) & \eql 289 x^8-1372 x^6+1302 x^4-1372 x^2+289\,,\\
P_{\Re\,z_3}(x) & \eql 1397 x^4-2380 x^3+4350 x^2-2380 x+245\,,\\
P_{\Im\,z_3}(x) & \eql 1951609 x^8-12150144 x^6+22045824 x^4-17915904 x^2+2985984\,.
\end{split}
\end{equation}

%%%%%%%%%%%%%%%%%%
\item[]
$\bf S1039230~\&~\overline S1039230$ \cite{Borghese:2013dja}
\begin{equation}\label{}
\begin{split}
z_1\eql z_2\eql \sqrt{5-2 \sqrt{6}}\approx 0.3178372 \,,\qquad z_3\eql -i\,\sqrt{2-\sqrt{3}}\approx -0.517638\rmi\,,
\end{split}
\end{equation}
\begin{equation}\label{}
\cals P\eql -6\sqrt 3 \approx -10.39230\,.
\end{equation}
\qquad
Symmetry: $\SO(4)\,,\quad \cals N=0\,.$\qquad Orbits: $\langle 1,g_H,g_R',g_Hg_R'\rangle \,.$
\begin{equation}\label{}
P_{\Re\,z_{1,2}}(x) \eql x^4-10 x^2+1\,,\qquad P_{\Im\,z_{3}}(x)\eql x^4-4 x^2+1\,.
\end{equation}

%%%%%%%%%%%%%%%%%
\item[]
$\bf S1384096~\&~\overline  S1384096$ \cite{Comsa:2019rcz}
\begin{equation}\label{S1384096}
\begin{split}
z_1 & \approx 0.1696360 + 0.1415740 \,i \,,\\
z_2 & \approx  0.4833214 + 0.3864058\,i \,,\\
z_3 & \approx -0.3162021 - 0.5162839 \,i\,, 
\end{split}
\end{equation}
\begin{equation}\label{}
\cals P\approx -13.840964\,.
\end{equation}
\qquad
Symmetry: $\SO(3)$\,,\quad $\cals N=1$\,.\qquad Orbit: $\langle 1,g_H,g_C,g_Hg_C\rangle$\,.  

\qquad Comment:
See Section~\ref{sec:newpoint}.
\\[- 6 pt]

%%%%%%%%%%%%%%%%%
\item[]
$\bf S1400000$ \cite{Warner:1983du}
\begin{equation}\label{eq:SO3SO3crit}
\begin{split}
z_1&=z_2 = {1\over 2}(1+\rmi)\sqrt{3-\sqrt 5} \approx 0.4370160(1+i) \,, \qquad  z_3=\bar{z}_3=0\,,\end{split}
\end{equation}
\begin{equation}\label{}
\qquad 
\mathcal{P} = -14\,.
\end{equation}
\qquad
Symmetry: $\SO(3)\times \SO(3)$\,,\quad $\cals N=0$\,.\qquad Orbit: $\langle 1,g_H,g_C,g_Hg_C\rangle$\,. 
\begin{equation}\label{}
P_{\Re\,{z_1,2}}(x)\eql P_{\Im\,{z_1,2}}(x)\eql 4 x^4-6 x^2+1\,.
\end{equation}

\qquad Comment: This point is non-supersymmetric, but  perturbatively stable \cite{Fischbacher:2010ec}.
\\[- 6 pt]
%%%%%%%%%%%%%%%%%%%%%%

\item[]
$\bf S1424025$ \cite{Fischbacher:2011jx}
\begin{equation}\label{}
\begin{split}
z_1 & \approx 0.4490422 + 0.4843455 \,i \,,\\
z_2 & \approx 0.3750597 + 0.2850151 \,i \,,\\
z_3 & \approx -0.04539020  \,,
\end{split}
\end{equation}
\begin{equation}\label{}
\cals P\approx -14.24026\,.
\end{equation}
\qquad
Symmetry: $\SO(3) \,,\quad \cals N=0$\,.\qquad Orbit: $\langle 1,g_H,g_C,g_Hg_C\rangle$\,. 

\vskip1ex

\qquad Comment: Coordinates are known algebraically, cf. Appendix~\ref{App:algebraicpositions}.

\vskip3ex

%%%%%%%%%%%%%%%%%%%%%
\noindent
$\bf S2096313~\&~\overline  S2096313$
\begin{equation}\label{}
\begin{split}
z_1 & \eql -i\,(2-\sqrt{5} )\approx 0.2360680\, i\,,\\
z_2 & \eql -\frac{i}{2}   (1-\sqrt{5})\approx 0.6180340\,i\,,\\
z_3 & \eql \frac{2}{41} \left(15-2 \sqrt{5}\right)-\frac{i}{41}\, \sqrt{21 \left(49-12 \sqrt{5}\right)}\approx
0.5135543- 0.5262366\,i\,,
\end{split}
\end{equation}
\begin{equation}\label{}
\cals P\eql -{75\over 8}\sqrt 5\approx -20.96314\,.
\end{equation}
\qquad Symmetry: $\rm SO(3)\times U(1)\,,\quad \cals N=0$\,.\qquad 
Orbits: $\langle 1,g_H\rangle$\,.
\begin{equation}\label{}
\begin{split}
P_{\Im\,z_1}(x)& \eql x^2-4 x-1\,,\qquad P_{\Im\,z_2}(x)\eql x^2-x-1\,,\\
P_{\Re\,z_3}(x)& \eql 41 x^2-60 x+20\,,\qquad P_{\Im\,z_3}(x)\eql 1681 x^4-2058 x^2+441\,.
\end{split}
\end{equation}

%%%%%%%%%%%%%%%%%%%%
\item[]
$\bf S2443607~\&~\overline  S2443607$ 
\begin{equation}\label{}
\begin{split}
z_1 & \approx 0.2187103 + 0.1800635  \,i\,,\\
z_2 & \approx -0.2046730 + 0.4973759  \,i\,,\\
z_3 & \eql 0.4188443 - 0.6668735 \,i\,,
\end{split}
\end{equation}
\begin{equation}\label{}
\cals P\approx -24.43607\,.
\end{equation}
\qquad Symmetry: $\rm SO(3)\,,\quad \cals N=0$\,.\qquad 
Orbits: $\langle 1,g_H,g_C,g_Hg_C\rangle$\,.

\vskip1ex

\qquad Comment: Coordinates are known algebraically, cf. Appendix~\ref{App:algebraicpositions}.

\vskip3ex
  
\end{itemize}

%%%%%%%%%%%%%%%%
\subsection{Subtruncations}
\label{ssec:subtr}
%%%%%%%%%%%%%%%%

The locations of the critical points above suggest a number of futher truncations to simpler subsectors. In particular we have the $\rm G_2$-invariant truncation, which in the parametrization used in \cite{Bobev:2013yra} is obtained by setting
\begin{equation}\label{eq:G2limit}
z_1=- z_2=-z_3=z .
\end{equation}
The superpotential and the K\"ahler potential reduce then to 
\begin{equation}\label{eq:G2suppot}
\mathcal{W}_{{\rm G}_2} = z^7+7z^4+7z^3+1\,,\qquad \mathcal{K}_{{\rm G}_2} = -7\log(1-z\bar{z}) \,.
\end{equation}
Within this truncation one finds 6 critical points: the $\SO(8)$ point, 
$\S0600000$, the $\SO(7)^+$ point,  $\S0668740$, the $\SO(7)^-$ points, $\S0698771$ and $\tt \overline S0698771$, and the $\rm G_2$ points, $\S0719157$ and $\tt\overline S0719157$.

Other simple truncations to one complex scalar field  are the $\SO(4)\times\SO(4)$-invariant truncation obtained by setting
 \begin{equation}\label{}
z_1=z_3=0\,,\qquad z_2\eql z\,,
\end{equation}
with 
\begin{equation}\label{eq:SO4suppot}
\mathcal{W}_{\SO(4)\times\SO(4)} =1 \,,\qquad \mathcal{K}_{\SO(4)\times\SO(4)} = -\log(1-z \bar{z})\,,
% \qquad \mathcal{P}_{\SO(4)\times\SO(4)} = -2\frac{3-z_2\bar{z}_2}{1-z_2\bar{z}_2} \,.
\end{equation}
and the  $\SU(3)\times \rU(1)^2$-invariant truncation
\begin{equation}\label{eq:SU3limit}
z_1=z_2=0\,, \qquad z_3=-z\,, 
\end{equation}
with
\begin{equation}\label{eq:SU3suppot}
\mathcal{W}_{\SU(3)\times \rU(1)^2} = z^3+1\,,\qquad \mathcal{K}_{\SU(3)\times \rU(1)^2} = -3\log(1-z\bar{z})\,.
%\quad \mathcal{P}_{\SU(3)\times \rU(1)^2} = -6\frac{1+\zeta\bar{\zeta}}{1-\zeta\bar{\zeta}} \,.
\end{equation}
Although there are no critical points other than the maximally supersymmetric $\S0600000$ one, those truncations admit nontrivial generalizations of the RG flows dual to three dimensional field theories with interfaces \cite{Bobev:2013yra}.

%%%%%%%%%%%%%%%%
\section{Numerical searches - an outline of the method}
\label{numerics}
%%%%%%%%%%%%%%%%

The TensorFlow code that was published alongside~\cite{Comsa:2019rcz} is readily adapted to search for critical points not on the full 70-dimensional scalar manifold but on submanifolds that are invariant under some residual symmetry, such as the six-dimensional space studied here.
As for the unconstrained problem, one starts from some random linear combination of the six $\rm E_{7(7)}$ generators that is sufficiently close to the origin for the numerical value of the potential to still be reliable, and then numerically minimizes the violation of the (un-truncated) stationarity-condition.

This way, one manages to discover all the critical points on the scalar manifold listed in Section~\ref{subsec:critpot} after about~$10\,000$ such iterations.
We observe that some tweaks to the code as published can improve search efficiency further.
In particular, it turns out to be beneficial to not use a second order numerical optimization method (such as BFGS) directly, but to first perform a few hundred gradient descent steps per iteration before switching to such a more advanced method.
Intuitively, if a second order optimization method gets to see from the start a sum of stationarity-violation contributions having very different scale, it will tend to be mostly sensitive to the most important contribution's second order approximation and hence in its first few steps move to very similar positions on the manifold, counteracting the need for good exploration.

Even with such tricks, the TensorFlow based search is fundamentally only a probabilistic method that converges to the various critical points with very uneven likelihood.
So, it might be conceivable that, even with much computational effort, some critical points remain undiscovered.
It hence makes sense to look for alternative approaches to the problem of finding critical points of algebraic functions (or, equivalently, intersections of algebraic varieties) on spaces of moderate dimension.

While simple techniques based on interval arithmetic or affine arithmetic appear too limited to conveniently study the restricted six-dimensional potential at hand, this problem is still well within reach of modern computational algebraic geometry.

For a task like this, one will typically want to first employ a modern computational algebraic geometry package such as~\cite{M2} to reduce/factorize the problem, and then use an adaptive-precision homotopy continuation solver such as Bertini2~\cite{BHSW06} that uses the algorithm described in~\cite{Hauenstein2017226} to systematically find solutions of the generalized problem with complex coordinates.
These solutions then have to be filtered, discarding all those with non-real coordinates.
Depending on the difficulty of the task, computations may take hours to days with Bertini2, and while this approach might hypothetically still miss some solutions, this is not observed to happen in practice.

These numerical algebraic geometry methods found the same list of critical points for our six-scalar model as the TensorFlow based search. We also note that for the eight-dimensional scalar manifold studied in~\cite{Bobev:2018uxk}, the same numerical methods manage to reproduce the list of critical points presented in that publication without uncovering additional ones\footnote{We thank Jonathan Hauenstein for performing these calculations.}.

For all critical points listed in Section~\ref{sec:otherpoints}, one can obtain algebraic expressions for their location, the potential, and other physical properties via inverse symbolic computation (i.e.\ employing the PSLQ algorithm). However, for $\S1424025$ and $\S2443607$, these expressions become rather lengthy.

%%%%%%%%%%%%%%%%%%%%%%%%%%%%%%%%%%%%%
\section{Holographic RG flows}
\label{sec:flows}
%%%%%%%%%%%%%%%%%%%%%%%%%%%%%%%%%%%%%

The truncation derived in Section~\ref{sec:truncation} makes it feasible to study explicitly supersymmetric holographic RG flows 
to the  new $\cals N=1$ critical point. For other supersymmetric critical points that lie within our truncation such flows have been constructed previously in  \cite{Ahn:2000mf,Ahn:2001kw, Ahn:2001by,Ahn:2002qga,Bobev:2009ms}.

The RG flows we are interested in are given  by domain wall solutions of the BPS equations in $\mathcal{N}=8$ $d=4$ supergravity with the metric of the form 
\begin{equation}\label{eq:4dDWAnsatz}
ds^2 = e^{2A(r)}ds^2_{1,2}+dr^2 \,,
\end{equation}
where $ds^2_{1,2}$ is the metric on the three-dimensional Minkowski space. Setting the supersymmetry variations of the supergravity fermion fields to zero,  the standard analysis, see e.g.\ \cite{Bobev:2009ms}, yields   the following system of  BPS equations: 
\begin{align}\label{eq:BPSeq}
z_a'(r)  = \mp\sqrt 2\,g\, e^{\cals K/2}\,\mathcal{K}^{a\bar{b}}\,{\cals W\over |\cals W|} & \nabla_{\bar{b}}\overline{\mathcal{W}}\,,\qquad 
\bar{z}_{\bar{b}}'(r)  = \mp\sqrt 2\,g\, e^{\cals K/2}\,\mathcal{K}^{a\bar{b}}\,{\overline{\cals W}\over |\cals W|} \nabla_{a}{\mathcal{W}}\,,\\[6 pt]
A'(r) &= \pm\sqrt{2}\, g\,|\cals W|\,,\label{eq:BPSA}
\end{align}
for the dependence of the scalar fields, $z_a$ and $\bar z_a$, and the metric function, $A$, on the radial coordinate.\footnote{As usual,  a prime denotes a derivative with respect to $r$.} In an $\cals N=1$ theory  these equations are completely determined by the K\"ahler potential and the superpotential of the truncated model as indeed they are in \eqref{eq:BPSeq} and \eqref{eq:BPSA}. The choice of sign in \eqref{eq:BPSeq} reflects the freedom to choose the sign of the radial coordinate $r$ in \eqref{eq:4dDWAnsatz}. In the calculations below we choose the upper sign in  \eqref{eq:BPSeq}.

It follows from the discussion in Sections~\ref{sec:newpoint} and \ref{sec:otherpoints} that
the system of ODEs \eqref{eq:BPSeq} has  critical points at the $\SO(8)$ vacuum \hyperref[so8point]{$\tt S0600000$}, the ${\rm G}_2$ vacua  \hyperref[eq:G2crit]{$\tt S0719157$}, the $\rm SU(3)\times U(1)$ vacuum \hyperref[su3u1pt]{$\tt S0779422$}, as well as the new $\mathcal{N}=1$  $\SO(3)$ vacuum \hyperref[S1384096]{$\tt S1384096$}.  The expectation is that there should be  a web of domain wall solutions corresponding to RG flows between the superconformal fixed points of the dual ABJM theory. Indeed, in a simpler setting that included the first three points only, families of such flows were constructed explicitly in \cite{Bobev:2009ms}.

%%%%%%%%%%%%%%%%%%%%%%%%%%%%%%%%%%%%%%%%%%%%%%%%%%
\begin{table}[t]
\begin{center}
\begin{tabular}{@{\extracolsep{8 pt}}c  c c c c c c }
\toprule
\noalign{\smallskip}
Point & \multicolumn{6}{c}{$\delta_\alpha$} \\
\noalign{\smallskip}
\midrule
\noalign{\smallskip}
$\tt S0600000$ & 1.00000 & 1.00000 & 1.00000 & 1.00000 & 1.00000 & 1.00000 \\
$\tt S0719157$ & {3.44949} & 1.40825 & 1.40825 & 0.591752 & 0.591752 & $-$1.44949 \\
$\tt S0779422$ & 3.56155 & 2.56155 & 1.33333 & 0.666667 & $-$0.561553 & $-$1.56155 \\
$\tt S1384096$ & 4.80254 & 3.71632 & 1.25126 &   0.748735 &   $-$1.71632 &   $-$2.80254 \\
\bottomrule
\end{tabular}
\caption{\label{tbl-assexp} Asymptotic exponents, $\delta_\alpha$, at the four supersymmetric critical points.  }
\end{center}
\end{table}
%%%%%%%%%%%%%%%%%%%%%%%%%%%%%%%%%%%%%%%%%%%%%%%%%%

To study these domain wall solutions it proves convenient to split the complex scalar fields, $z_a$, into their real and imaginary parts, $z_a=x_a+i\,y_a$, $a=1,2,3$. At each critical point, the real fields, $(x_1,y_1)$, $(x_2,y_2)$ and $(x_3,y_3)$, collectively denoted by $\phi_\alpha$, $\alpha=1,\ldots,6$, have the asymptotic expansions
\begin{equation}\label{}
\phi_\alpha(r)\eql \sum_{\beta=1}^6 A_{\alpha\beta}\,e^{-\delta_\beta r/L}+\ldots\,,\qquad L^2\eql -{3\over\cals P}\,,
\end{equation}
where $L$ is the radius of the corresponding AdS$_4$ solution determined by the value of the potential $\mathcal{P}$ at a given critical point. The exponents $\delta_\alpha$ are related to the scaling dimensions, $\Delta_\alpha$, of the dual  operators by 
\begin{equation}\label{}
\delta_\alpha\eql \Delta_\alpha\qquad \text{or}\qquad \delta_\alpha\eql 3-\Delta_\alpha\,.
\end{equation}

\begin{figure}[H]
\begin{center}
\includegraphics[width=1.95in]{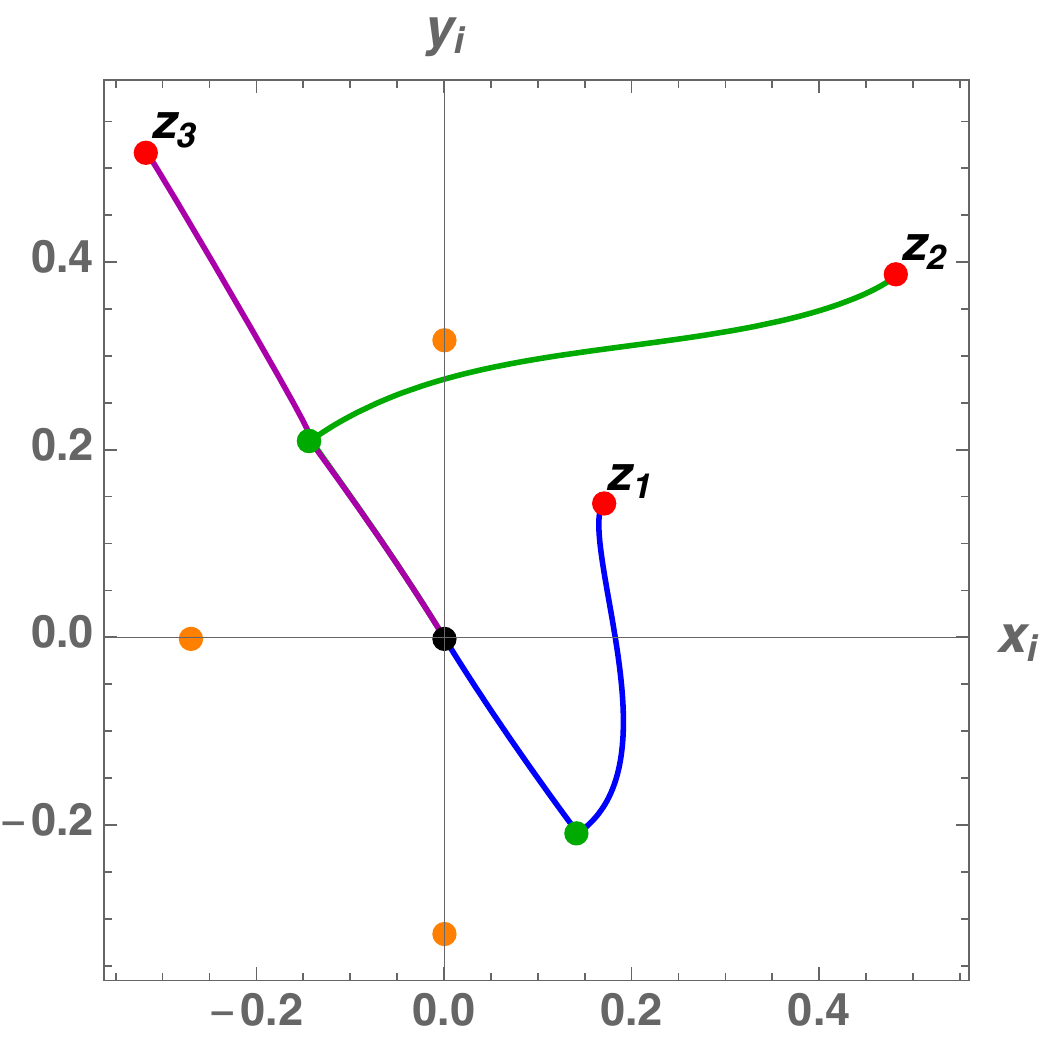}
\quad \includegraphics[width=1.95in]{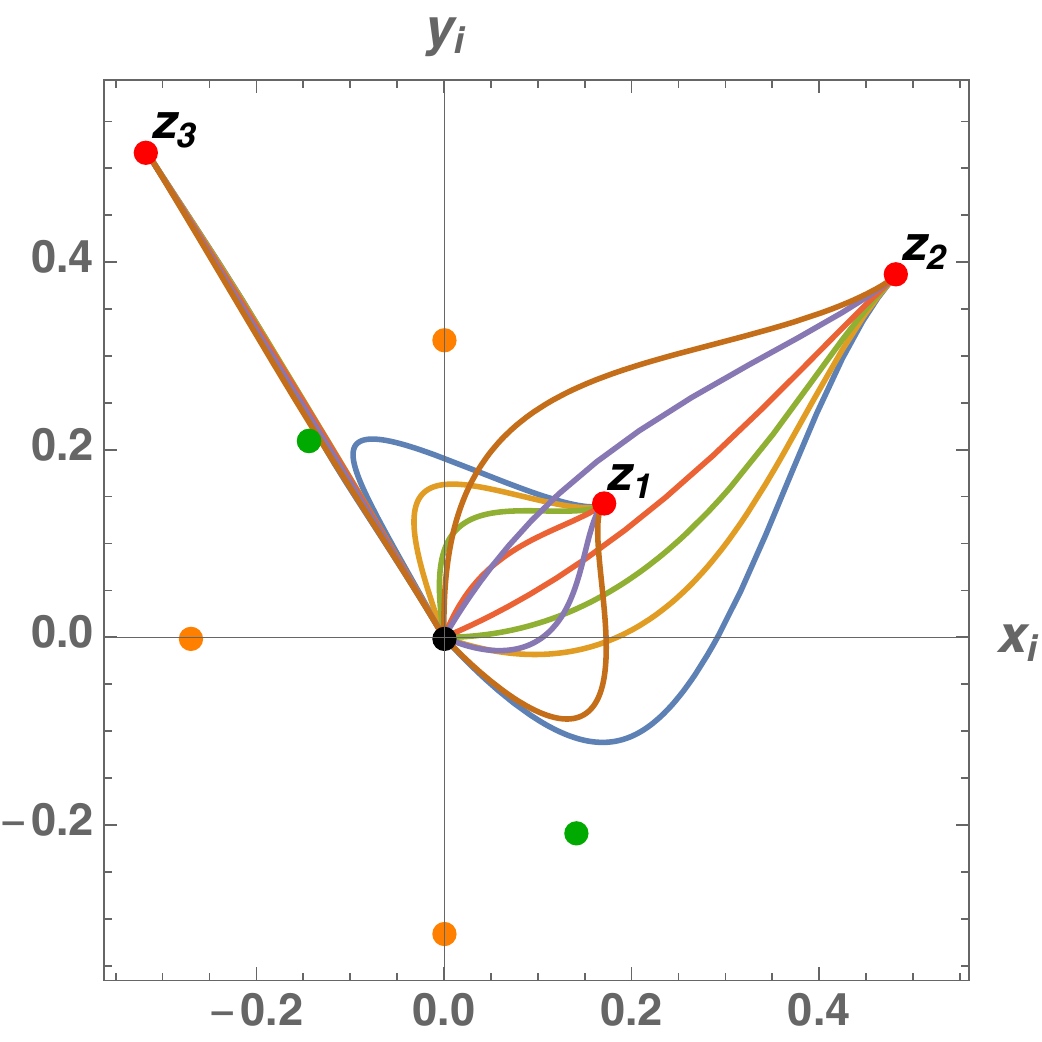}
\quad\includegraphics[width=1.95in]{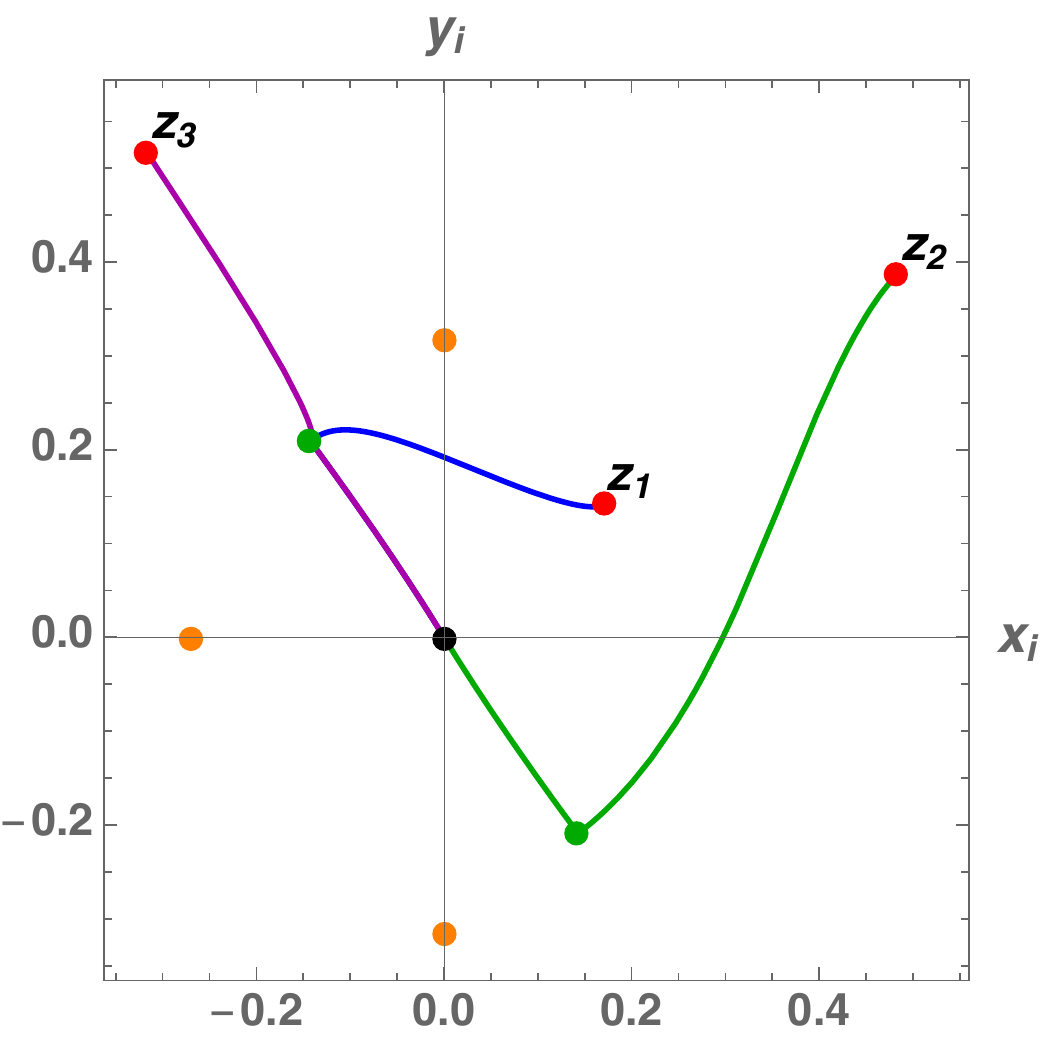}\\[12 pt]
\includegraphics[height=1.45in]{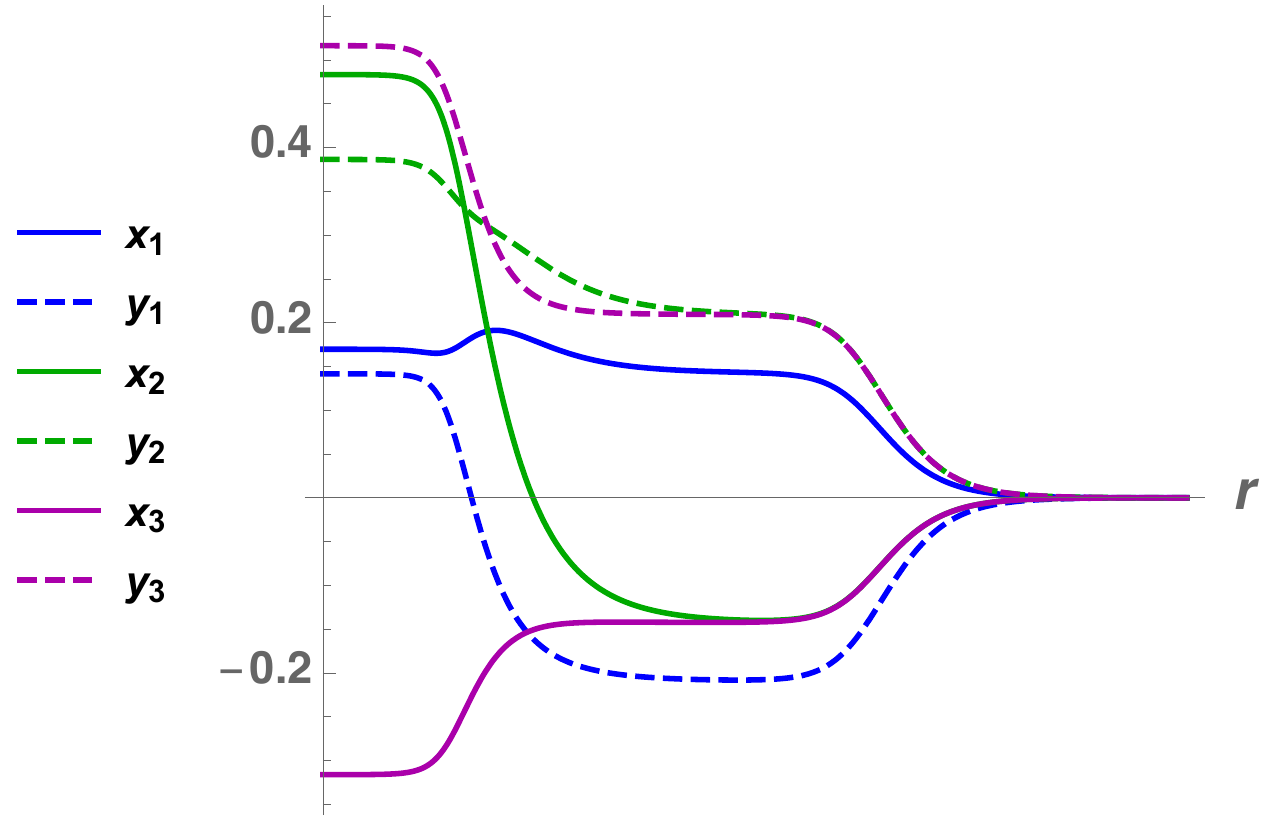}\quad \includegraphics[height=1.45in]{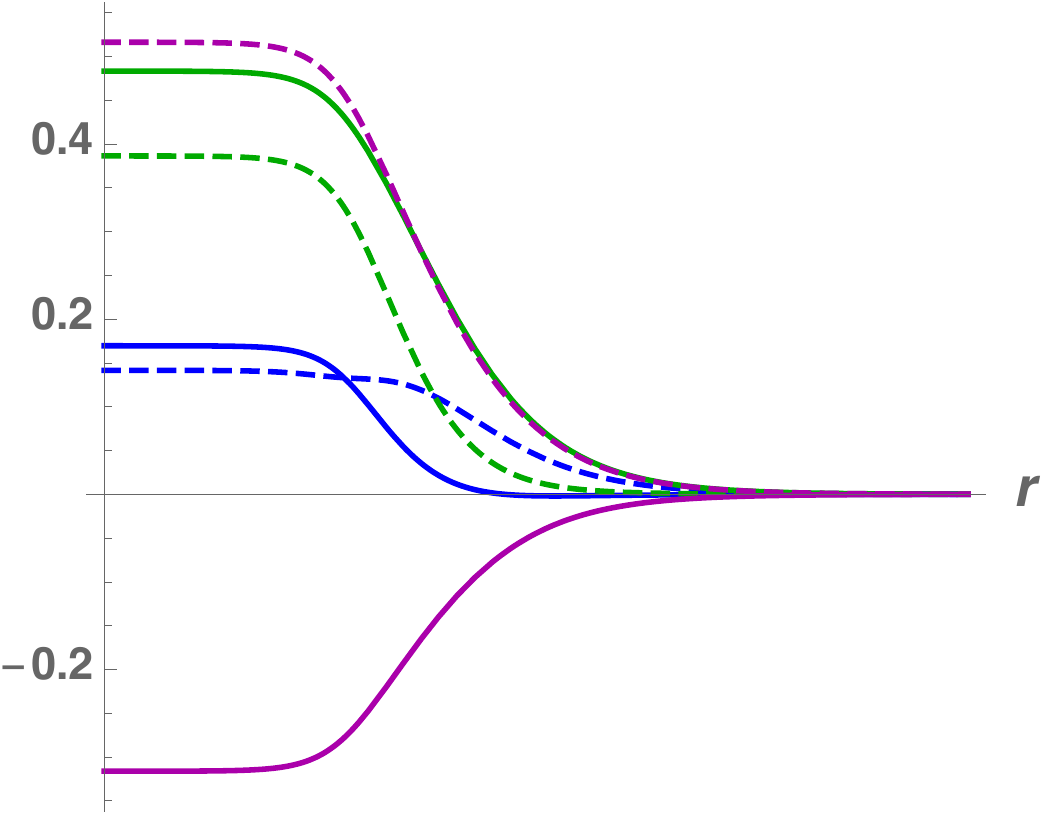}\quad
\includegraphics[height=1.45in]{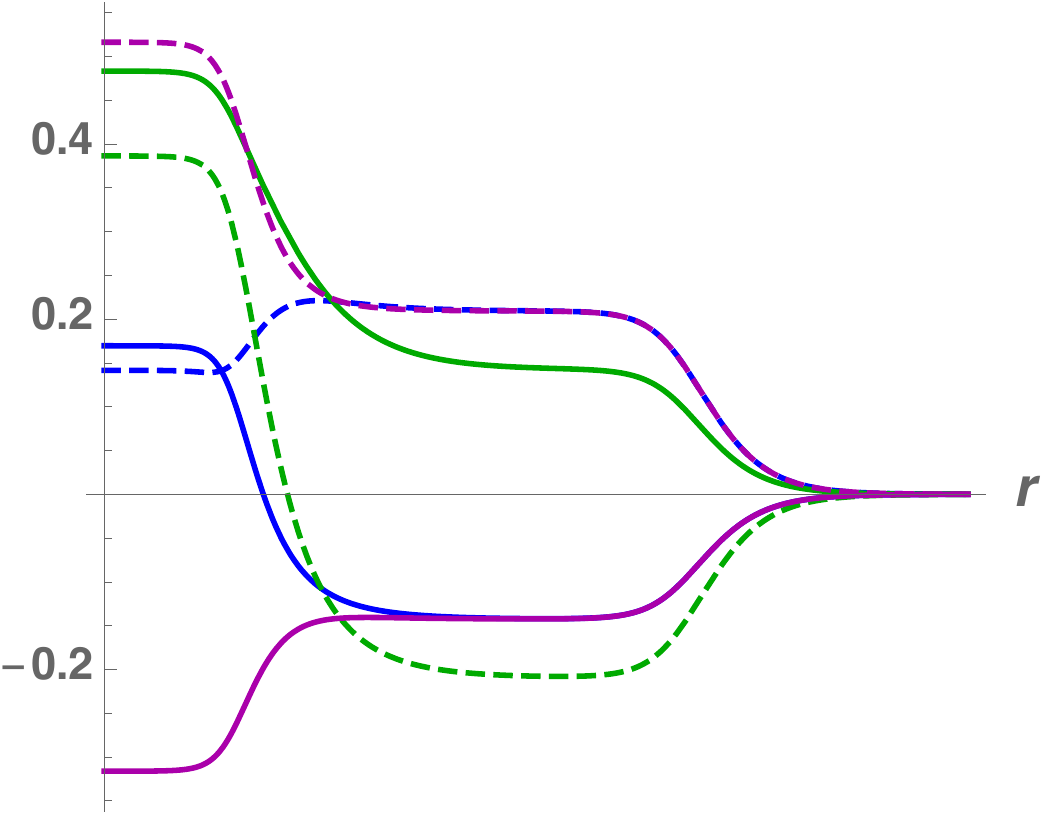}
\\[12 pt]
\includegraphics[height=1.25in]{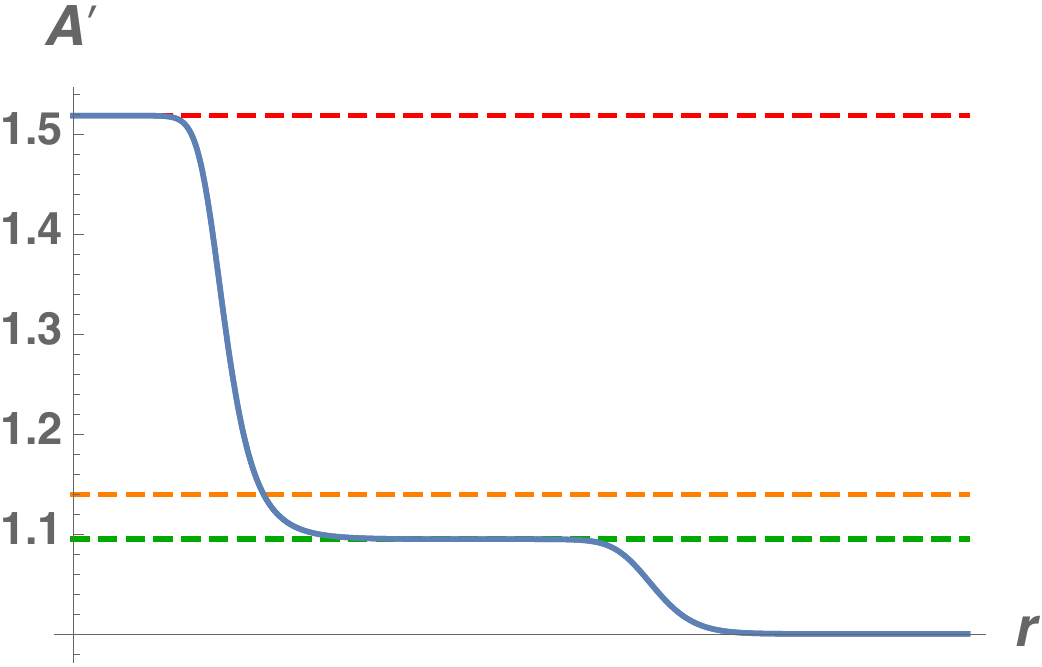}\quad \includegraphics[height=1.25in]{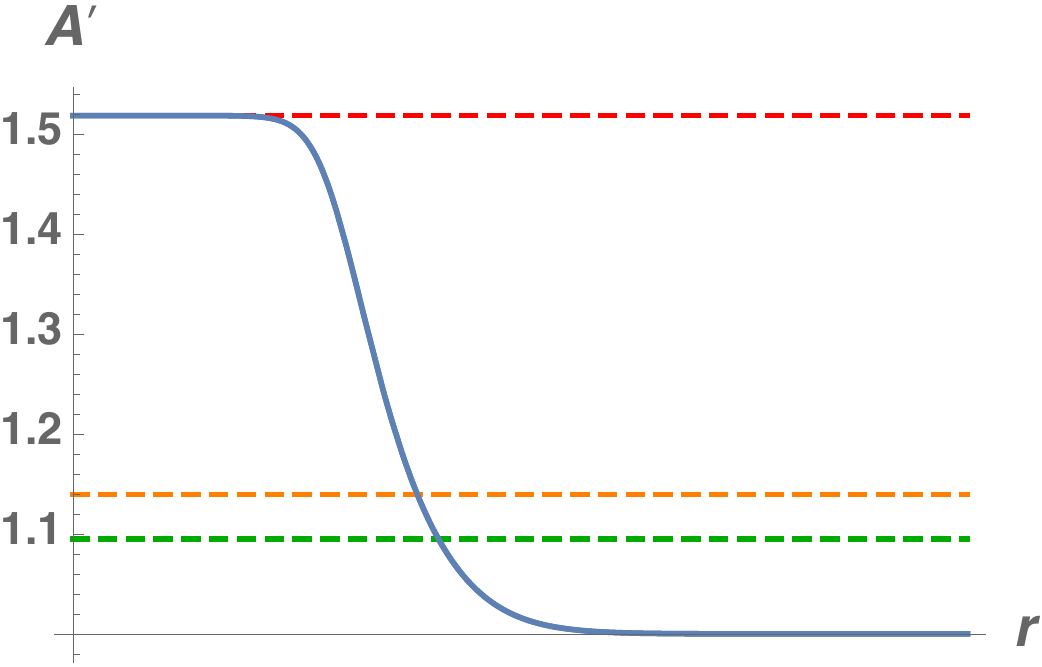}\quad \includegraphics[height=1.25in]{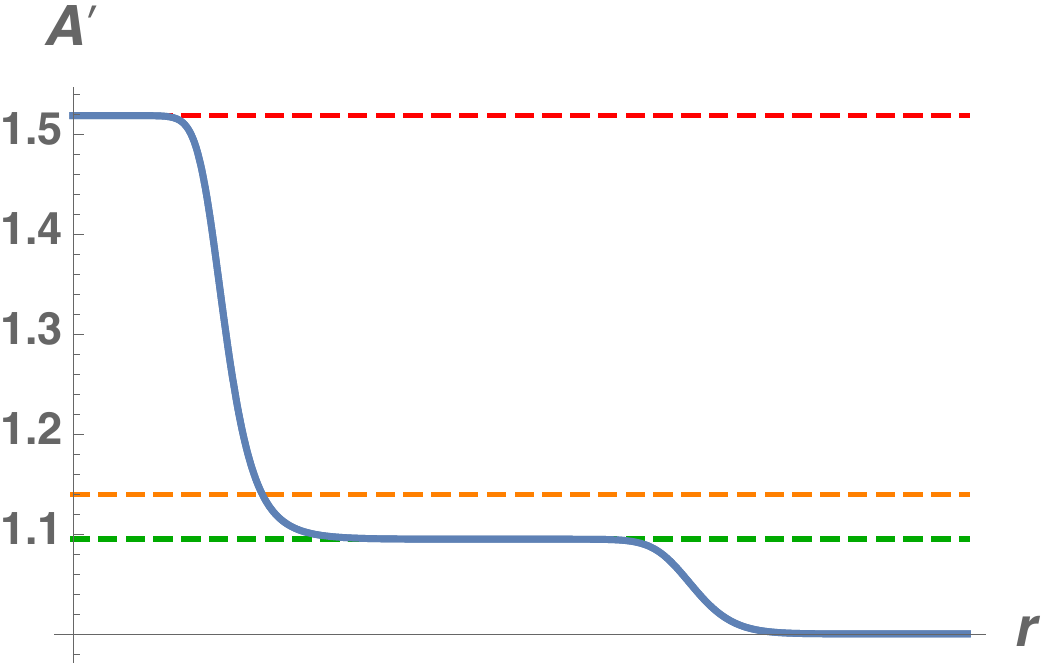}
\end{center}
\caption{\label{fig-so3so8} Numerical solutions to the BPS equations for RG flows from the $\SO(8)$ point to the $\SO(3)$ point. The generic flows (middle column) asymptote to flows through the $\rm G_2$ point (side columns). The top row shows the flows in the superimposed $z_1$, $z_2$ and $z_3$-planes. The colored dots represent the supersymmetric critical points: SO(8) (black), $\rm G_2$ (green), $\rm SU(3)\times U(1)$ (orange), and $\SO(3)$ (red). The middle row gives the radial dependence of the real scalars, $x_1,\ldots, y_3$. The bottom row gives $A'$ along the flows, which asymptotes to a constant that depends on the radius of the AdS$_4$ vacuum.}
\end{figure}

\begin{figure}[h]
\begin{center}
\includegraphics[height=1.4in]{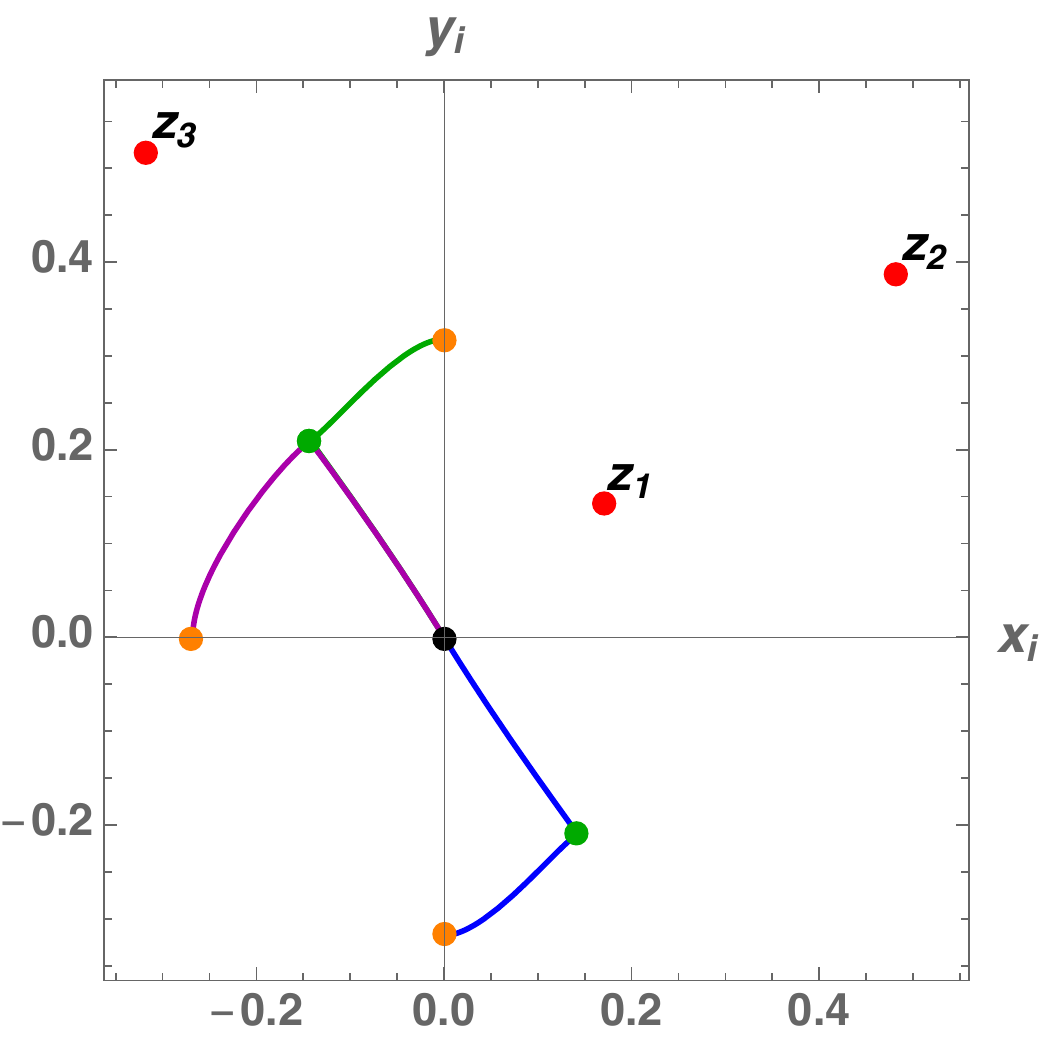}\quad \includegraphics[height=1.3in]{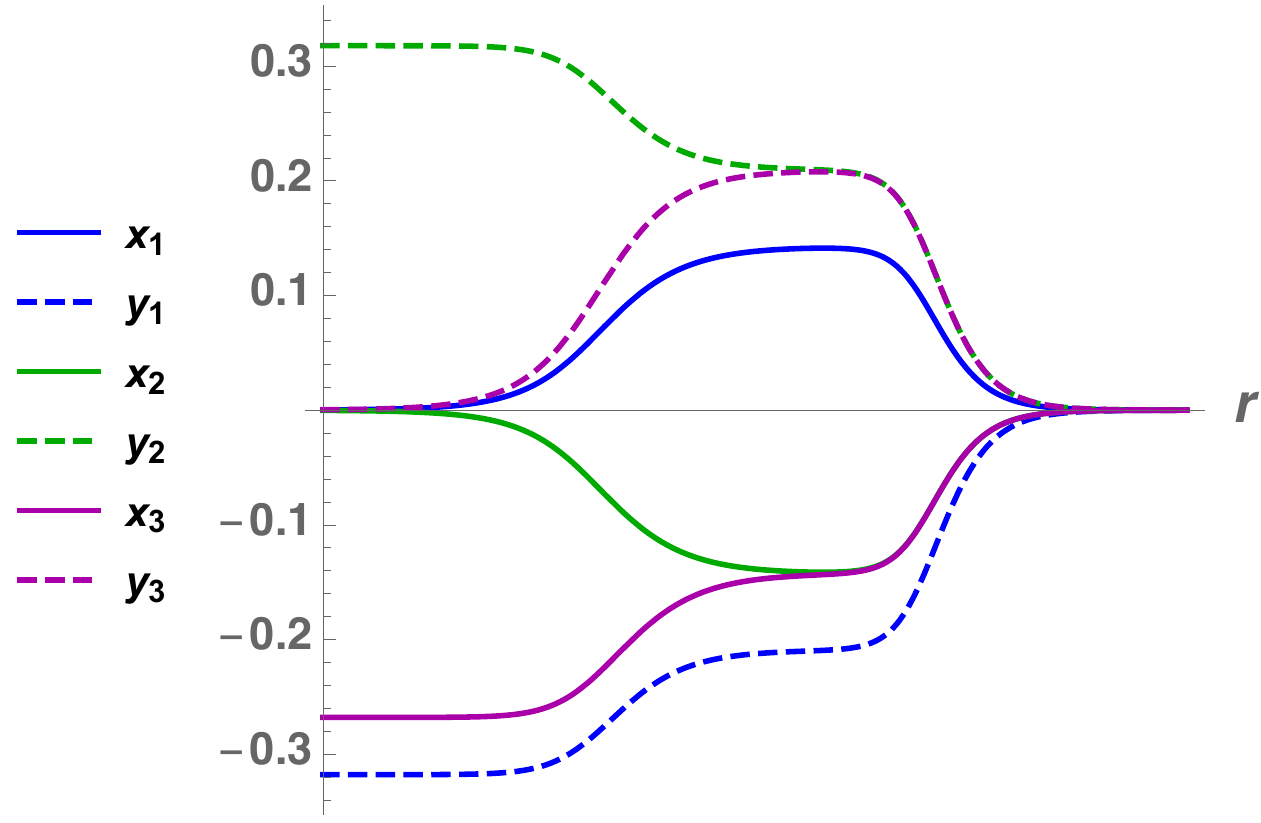}\quad
\includegraphics[height=1.2in]{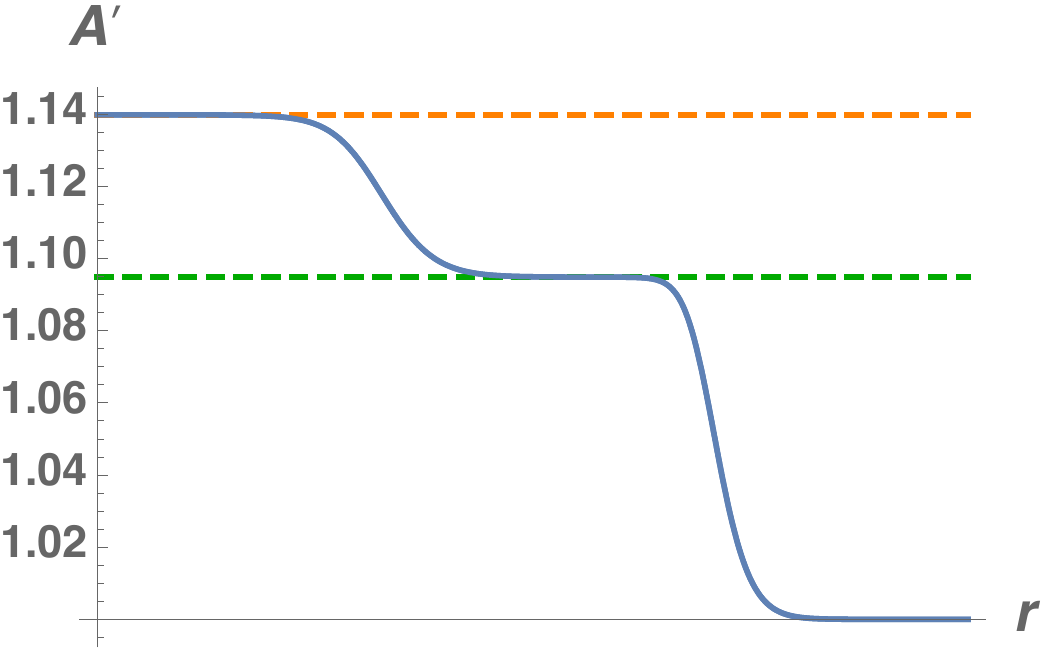}
\end{center}
\caption{\label{fig-su3g2so8} The RG-flow from the $\SO(8)$ point  to the $\SU(3)\times {\rm U}(1)$ point asymptotic to the $\rm G_2$ point.}
\end{figure}

The BPS equations   \eqref{eq:BPSeq} can be integrated numerically and, as expected, we find families of domain wall solutions interpolating between the $\SO(8)$ vacuum  and the new $\SO(3)$ vacuum. Examples of such solutions are shown in the middle column in Figure~\ref{fig-so3so8}. There are also finely tuned  solutions that realize holographically a ``triangular RG flow'' starting from the $\SO(8)$ vacuum  in the UV, approaching one of the two ${\rm G}_2$ vacua  and then ultimately ending in the $\SO(3)$  vacuum in the deep IR. Plots of those  flows are  shown in the  left and right columns in Figure~\ref{fig-so3so8}. Similarly, there are supersymmetric triangular RG flows, see   Figure~\ref{fig-su3g2so8}, interpolating between the $\SO(8)$, the $\rm SU(3)\times U(1)$, and the ${\rm G}_2$ critical points. Those solutions were first studied in \cite{Bobev:2009ms} and are also present within the consistent truncation  here.

However, using a simple shooting method we were not able to find  similar triangular RG flows involving the $\SO(8)$, the $\rm SU(3)\times U(1)$, and the $\SO(3)$ points or for that matter all four supersymmetric  points. Indeed, a more exhaustive numerical search using Machine Learning, to be discussed elsewhere, strongly suggests that such RG flows do not exist within our 6-scalar $\SO(3)\times \mathbb{Z}_2$-invariant truncation. Still, we suspect that those flows might exist within  a larger truncation, perhaps the $\ZZ_2\times \ZZ_2\times \ZZ_2$-invariant truncation discussed in  Section~\ref{sec:Z23}, whose remarkable properties make it a compelling candidate to look at.

%%%%%%%%%%%%%%%%%%%%%%%%%%%%%%%%%
%%%%%%%%%%%%%%%%%%%%%%%%%%%%%%%%%

To interpret this web of RG flows in the dual ABJM SCFT, it is convenient to employ the $\mathcal{N}=1$ superspace language and, using the same notation as in \cite{Bobev:2009ms}, consider  the following combinations of the eight chiral superfields,
\begin{equation}
\tilde{Z}_{a} = \Phi_{2a-1} +\rmi \Phi_{2a}\,, \qquad a=1,2,3,4\,.
\end{equation}
%

%%%%%%%%%%%%%%%%%%%%%%%%%%%%%%%%%
%%%%%%%%%%%%%%%%%%%%%%%%%%%%%%%%%

\noindent
Then the deformation of the ABJM superpotential,
\begin{equation}
\Delta W = \frac{1}{2}m_3\big(\tilde{Z}_1^2+\tilde{Z}_2^2+\tilde{Z}_3^2\big) + \frac{1}{2}m_7 \Phi_7^2+ \frac{1}{2}m_8 \Phi_8^2\,.
\end{equation}
breaks the conformal invariance of the ABJM theory, but preserves $\mathcal{N}=1$ supersymmetry. For general values of the mass parameters $m_3$, $m_7$ and $m_8$, the deformation preserves the $\SO(3)$ symmetry. The structure of the  holographic RG flows  above suggests that the IR dynamics of this model is controlled by a new  interacting $\mathcal{N}=1$ SCFT, which is the field theory dual of the new  $\mathcal{N}=1$ vacuum $\tt S1384096$. Note that for $m_3=0$ we recover the RG flows to the ${\rm G}_2$ and $\SU(3)\times \rU(1)$ critical points discussed in \cite{Bobev:2009ms}.

%%%%%%%%%%%%%%%%%%%%%%
\section{A $\ZZ_2\times \ZZ_2\times \ZZ_2$-invariant truncation}
\label{sec:Z23}
%%%%%%%%%%%%%%%%%%%%%

A lesson one should draw from the construction of the critical point
$\tt S1384096$ above, as well as from a similar construction
in \cite{Fischbacher:2010ec} of the $\cals N=1$ supersymmetric
critical point, $\tt S1200000$, which is not captured by our
truncation, is that discrete symmetries may lead to simple, explicit
truncations of the $\cals N=8$ supergravity that are accessible
analytically. In both of these constructions, the scalar manifold of
the truncated theory is simply a product of three Poincar\'e disks,
$\SU(1,1)/\rm  U(1)$. One may note that the same coset arises in the
so-called STU-model  with a particularly simple
superpotential \cite{Freedman:2013ryh}. Observing that in our construction, two of the three $\SU(1,1)$ factors are embedded non-regularly into $\rm E_{7(7)}$, it seems suggestive to try interpreting  the subspace studied here as originating from a collapsing of roots, starting from the maximal number of regularly embedded $\SU(1,1)$'s.  Both the enlargement of $\rm SU(1,1)\times SU(1,1)$ to $\rm G_{2(2)}$ when dropping the~$\ZZ_2$ symmetry (which is obtained from~$\rm SO(8)$ by collapsing three roots)
  as well as the coefficients in the K\"ahler potential~\eqref{eq:cKdef} suggest that one should look for a regular embedding of $\SU(1,1)^{\times 7}$ into $\rm E_{7(7)}$. The remarkable properties of this subgroup have been discussed in the context of qubit entanglement and black holes, cf.~\cite{Duff:2006ue} and subsequent research,  with a comprehensive review in~\cite{Borsten:2012fx},
 ``curious supergravities''    \cite{Duff:2010vy} and cosmology \cite{Ferrara:2016fwe}.

It turns out that  an $\cals N=1$ supersymmetric truncation with the scalar manifold given by the product of 7 Poincar\'e disks,
\begin{equation}\label{}
\left[\,{\rm SU(1,1)\over U(1)}\,\right]^7\,,
\end{equation}
 can be obtained using a discrete $\ZZ_2\times \ZZ_2\times \ZZ_2\subset \SO(8)$ symmetry. It can be constructed explicitly as follows.

Consider the   ${\rm S}\equiv \ZZ_2\times \ZZ_2\times \ZZ_2$ group generated by the following $\SO(8)$ rotations:
\begin{equation}\label{gZ23}
\begin{split}
g_1 & \eql \diag\left({1,1, 1, 1, -1, -1, -1, -1}\right)\,,\\
 g_2 & \eql\diag\left({1,1,-1,-1,1,1,-1,-1}\right)\,,\\
  g_3 & \eql\diag\left({1,-1,1,-1,1,-1,1,-1}\right)\,, 
 \end{split}
\end{equation}
in $\bfs 8_v$. The non-identity elements of this group  are naturally labelled by the 7 points on the Fano plane (see, e.g., \cite{Baez:2001dm})
\begin{equation}\label{u17gen}
h_1\eql g_1,\,,\quad h_2\eql g_2\,,\quad h_3\eql g_3\,,\quad h_4\eql g_1g_2\,,\quad h_5\eql g_2g_3\,\quad h_6\eql g_3g_1\,,\quad h_7\eql g_1g_2g_3\,,
\end{equation}
such that the product along each line is the identity.  It is straightforward to verify explicitly that the Lie subalgebra of $\frak e_{7(7)}$  invariant under $\rm S$ is precisely $\su(1,1)^{\oplus 7}$, with the compact generators, $\frak h_a$, of $\frak u(1)^{\oplus 7}$ given  by the matrices $\frak h_a=i\,h_a$, $a=1,\ldots,7$ in \eqref{u17gen}. Note that the generators $\fh_j$ are orthogonal to $\so(8)$ in $\su(8)$. Since each $\su(1,1)$ corresponds to a complex scalar field in the truncation, there is a natural identification of the resulting 7 complex scalars, $\zeta_a$,  with the points on the Fano plane, 
\begin{equation}\label{}
\zeta_a\quad \longleftrightarrow\quad \fh_a\eql i\,h_a\,,\qquad a=1,\ldots,7\,.
\end{equation}

As in Section~\ref{sec:truncation}, the superpotential can be read off from the eigenvalue of the $A^1_{ij}$ tensor along the unbroken supersymmetry. 
A direct calculation yields the result
\begin{equation}\label{}
\begin{split}
\cals W_{\ZZ_2^3} & \eql \zeta _1 \zeta _2 \zeta _3 \zeta _4 \zeta _5 \zeta _6 \zeta _7\\
&\quad +\zeta _1 \zeta _2 \zeta _3 \zeta _7+\zeta _1 \zeta _2 \zeta _5 \zeta _6+\zeta _1 \zeta _3 \zeta _4 \zeta _5+\zeta _1 \zeta _4 \zeta _6 \zeta _7+\zeta _2 \zeta _3 \zeta _4 \zeta _6+\zeta _2 \zeta _4 \zeta _5 \zeta _7+\zeta _3 \zeta _5 \zeta _6 \zeta _7\\
&\quad +\zeta _1 \zeta _2 \zeta _4+\zeta _1 \zeta _3 \zeta _6+\zeta _1 \zeta _5 \zeta _7+\zeta _2 \zeta _6 \zeta _7+\zeta _2 \zeta _3 \zeta _5+\zeta _3 \zeta _4 \zeta _7+\zeta _4 \zeta _5 \zeta _6+ 1\,,
\end{split}
\end{equation}
where each cubic term, together with the  complementary quartic term,\footnote{Complementary terms are defined by their product given by $\zeta_1\zeta_2\ldots\zeta_7$.}  corresponds to one of the 7
lines in the Fano plane. Even more remarkably, the terms in this polynomial match the 16 code words in the single-error-correcting (7,4) Hamming code \cite{hamming1950error}.

From the kinetic terms, or equivalently from the same embedding indices of $\su(1,1)$'s in $\frak e_{7(7)}$, we find the canonical K\"ahler potential
\begin{equation}\label{Kpot7}
\cals K\eql -\sum_{a=1}^7 \log(1-\zeta_a\bar\zeta_a)\,.
\end{equation}
This completely specifies the truncation with the scalar potential in this sector given by \eqref{eq:cPdef}.

One can check that all truncations of interest with fewer $\rm SU(1,1)/U(1)$ factors can be obtained by imposing additional continuous symmetry with respect to some subgroup of $\SO(8)$. This amounts to setting some scalars equal (up to a sign) and/or setting them to zero. For example, the truncation discussed in this paper can obtained by setting
\begin{equation}\label{}
\zeta_1\eql -z_2\,,\qquad \zeta_2\eql\zeta_6\eql\zeta_7\eql -z_3\,,\qquad 
\zeta_3\eql \zeta_4\eql \zeta_5  \eql z_1\,,
\end{equation}
upon which $\cals W_{\ZZ_2^3}$ reduces to \eqref{eq:cWdef} and the Kahler potential \eqref{Kpot7} to \eqref{eq:cKdef}. Similarly, the $\SO(2)\times \SO(2)\times \ZZ_2\times \ZZ_2$-invariant truncation  in \cite{Fischbacher:2010ec} is obtained by setting
\begin{equation}\label{}
\zeta_1\eql \zeta_3\eql i\,\xi_1\,,\qquad \zeta_2\eql \zeta_7\eql 0\,,\qquad \zeta_4\eql\zeta_5\eql \xi_2\,,\qquad \zeta_6\eql\xi_0\,,
\end{equation}
where $\xi_i$ are the scalar fields in \cite{Fischbacher:2010ec}. Finally, the superpotential of the STU model is obtained by keeping just one cubic term, for example by setting
\begin{equation}\label{}
\zeta_3\eql\zeta_5\eql\zeta_6\eql\zeta_7\eql 0\,.
\end{equation}

A preliminary numerical search has revealed~48 critical points which, as expected, include all 5 supersymmetric points: $\tt S0600000$, $\tt S0719157$, $\tt S0779422$, $\tt S1200000$, $\tt S1384096$,  and the non-supersymmetric stable point, $\tt S1400000$.
It is clear that this truncation should be the natural arena to study the holographic RG flows between the supersymmetric critical points and to look for the interplay between the structure of the critical points and the underlying octonion structure in the truncation. For further details we refer the reader to the follow up publication \cite{inprogress}.

%%%%%%%%%%%%%%%%%%%%%%%%%%%%%%%%%%%%%
\section{Conclusions}
\label{sec:conclusions}
%%%%%%%%%%%%%%%%%%%%%%%%%%%%%%%%%%%%%

In this paper we presented an explicit construction of a new AdS$_4$ vacuum of $\mathcal{N}=8$ supergravity which preserves $\mathcal{N}=1$ supersymmetry. We also discussed the spectrum of all supergravity fields around this vacuum and its relation to operators in the holographically dual CFT. Moreover, we constructed numerical holographic RG flow solutions which interpolate between this new vacuum and other supersymmetric vacua of the $\mathcal{N}=8$ supergravity.

One interesting outcome of combining  an explicit analytic truncation with the numerical methods using TensorFlow or Bertini2 is a discovery of two additional non-supersymmetric and  perturbatively unstable AdS$_4$ vacua that were not identified in the numerical search in \cite{Comsa:2019rcz}. A possible explanation  is that the numerical algorithms in \cite{Comsa:2019rcz} were applied to the full 70-dimensional scalar manifold of the $\mathcal{N}=8$ supergravity whereas here the search could be restricted to a  simpler and explicitly known potential that depends on  6 scalars only. This also points to a possible strategy for refining the numerical search by restricting it to  scalar submanifolds that are invariant under  continuous and/or discrete symmetries of  the points that had been found already.

Out of all known AdS$_4$ vacua in the $\mathcal{N}=8$ supergravity there are only 6 that are perturbatively stable. The $\SO(3)\times \mathbb{Z}_2$ supergravity truncation discussed in this paper contains 5 of these vacua. The one not included is the $\rU(1)\times \rU(1)$ $\mathcal{N}=1$ vacuum studied in \cite{Fischbacher:2010ki,Fischbacher:2010ec}. The larger truncation with 14 scalar fields presented in Section~\ref{sec:Z23} contains all 6 perturbatively stable AdS$_4$ vacua and therefore is a natural starting point for the study of explicit holographic RG flows between them. In fact, by imposing additional $\rm U(1)$ symmetry, one may further truncate to 10 scalar fields, while preserving all the interesting critical points. At the end such an analysis  will elucidate the phase structure of the ABJM SCFT and will provide a rich testing ground for the ``$\mu$-theorem'' discussed in \cite{Gukov:2015qea}.

The results presented here and in \cite{Comsa:2019rcz} suggest that one should apply similar techniques  to investigate the vacuum structure of other maximal supergravity theories using an amalgam of analytic and numerical methods. Two particularly interesting examples which can be embedded in string theory and have well-understood holographic duals are  the $\mathcal{N}=8$ ${\rm ISO}(7)$ gauged four-dimensional supergravity \cite{DallAgata:2011aa,DallAgata:2014tph,Guarino:2015qaa} and the maximal five-dimensional $\SO(6)$ gauged supergravity. We expect to report some preliminary results shortly \cite{BFGP}.

%%%%%%%%%%%%%%%%%%%%%%%%%%%%%%%%%%%%%
\bigskip
\bigskip
\leftline{\bf Acknowledgements}
\smallskip
\noindent 
N.B.\ and K.P.\  are grateful to Fri\dh rik Freyr Gautason and  Silviu Pufu for interesting discussions. T.F.\ would like to thank Jyrki Alakuijala and Rahul Sukthankar for feedback and encouragement on this work, Moritz Firsching and Sameer Agarwal for useful discussions on homotopy continuation methods, and Jonathan Hauenstein for confirming TensorFlow results with Bertini2. We also thank Moritz Firsching for providing minimal polynomials for the complex coordinates of the two algebraically most challenging critical points. The work of NB is supported in part by an Odysseus grant G0F9516N from the FWO and by the KU Leuven C1 grant ZKD1118 C16/16/005. KP is supported in part by DOE grant DE-SC0011687. NB and KP are grateful to the Mainz Institute for Theoretical Physics (MITP) of the DFG Cluster of Excellence PRISMA$^+$ (Project ID 39083149), for its hospitality and its partial support during the initial stages of this project.

%%%%%%%%%%%%%%%%%%%%%%%%%%%%%%%%%%%%%
\appendix
\section{Some group theory}

\label{appendixA}
%%%%%%%%%%%%%%%%%%%%%%%%%%%%%%%%%%%%%

We use the convention in which the gravitino, $\psi_\mu{}^i$, transforms in  $\bfs 8_v$, the spin-1/2 fermions are in the $\bfs{56}_v$ while the scalars and the pseudoscalars in ${\bf 35}^+={\bf 35}_s$ and ${\bf 35}^-={\bf 35}_c$ representations of $\so(8)$. The metric is neutral under $\so(8)$ and the gauge field is in the adjoint.

The commutant of  the $\so(3)\simeq \su(2)$ symmetry algebra in $\so(8)$ is $\u(1)\times \u(1)$. It arises from the following chain of maximal subalgebras:
\begin{equation}\label{}
\so(8)\supset \su(4)\times \u(1)_2\supset \su(3)\times \u(1)_1\times \u(1)_2\supset \su(2)\times  \u(1)_1\times \u(1)_2\,.
\end{equation}

The corresponding branchings of the $\so(8)$ representations relevant for our analysis are as follows:\footnote{We use the same group theory conventions as in \cite{Yamatsu:2015npn}.}
\begin{equation}\label{eq:8vbranch}
\begin{split}
\bfs{8}_v\quad &\longrightarrow\quad \bfs{4}_1+\bar{\bfs 4}_{-1} \quad\longrightarrow\quad \bfs{3}_{1,1}+\bfs{1}_{-3,1}+\bar{\bfs 3}_{-1,-1}+\bfs{1}_{3,-1}\\
\quad &\longrightarrow \quad \bfs{3}_{1,1}+\bfs{1}_{-3,1}+\bar{\bfs 3}_{-1,-1}+\bfs{1}_{3,-1}\,.
\end{split}
\end{equation}
For the vectors, the branching is
\begin{equation}\label{eq:28branch}
\begin{split}
\bfs{28} \quad\longrightarrow\quad & {\bf15}_{0} + \bfs{6}_{2}+ \bfs{6}_{-2}+ \bfs{1}_{0}\\
 \quad\longrightarrow\quad &\bfs{8}_{0,0}+\bfs{3}_{4,0}+\bar{\bfs 3}_{-4,0}+\bfs{1}_{0,0}+\bfs{3}_{-2,2}+\bar{\bfs 3}_{2,2}+\bfs{3}_{-2,-2}+\bar{\bfs 3}_{2,-2}+\bfs{1}_{0,0}\\
 \quad\longrightarrow\quad &\bfs{5}_{0,0}+\bfs{3}_{0,0}+\bfs{3}_{4,0}+\bfs{3}_{-4,0}+\bfs{1}_{0,0} +\bfs{3}_{-2,2}+\bfs{3}_{2,2}+\bfs{3}_{-2,-2}+\bfs{3}_{2,-2}+\bfs{1}_{0,0}\,.\\
\end{split}
\end{equation}
For the scalars we have
\begin{equation}\label{eq:35sbranch}
\begin{split}
\bfs{35}_s\quad\longrightarrow\quad &\bfs{20}'_{0}+ \bfs{6}_{2} + \bfs{6}_{-2} + \bfs{1}_{4} + \bfs{1}_{0} + \bfs{1}_{-4}\\
\longrightarrow\quad & \bar{\bfs 6}_{-4,0}+\bfs{8}_{0,0}+\bfs{6}_{4,0}+\bfs{3}_{-2,2}+\bar{\bfs 3}_{2,2}+\bfs{3}_{-2,-2}+\bar{\bfs 3}_{2,-2}+\bfs{1}_{0,4}+\bfs{1}_{0,0}+\bfs{1}_{0,-4}\\
\longrightarrow\quad & \bfs{5}_{-4,0}+\bfs{1}_{-4,0}+\bfs{5}_{0,0}+\bfs{3}_{0,0}+\bfs{5}_{4,0}+\bfs{1}_{4,0}+\bfs{3}_{-2,2}+\bfs{3}_{2,2}\\
&\qquad\qquad\qquad\qquad\qquad\qquad\qquad+\bfs{3}_{-2,-2}+\bfs{3}_{2,-2}+\bfs{1}_{0,4}+\bfs{1}_{0,0}+\bfs{1}_{0,-4}\,,
\end{split}
\end{equation}
and for the pseudoscalars we find
\begin{equation}\label{eq:35cbranch}
\begin{split}
\bfs{35}_c\quad\longrightarrow\quad & \bfs{10}_{-2} + \bfs{15}_{0} + \overline{\bfs{10}}_{2}\\
\quad\longrightarrow\quad & \bar{\bfs 6}_{2,-2}+\bfs{3}_{-2,-2}+\bfs{1}_{-6,-2}+ \bfs{8}_{0,0}+\bfs{3}_{4,0}+\bar{\bfs 3}_{-4,0}+\bfs{1}_{0,0} + \bfs{6}_{-2,2}+\bar{\bfs 3}_{2, 2}+\bfs{1}_{6, 2}\\
\quad\longrightarrow\quad & \bfs{5}_{2,-2}+\bfs{1}_{2,-2}+\bfs{3}_{-2,-2}+\bfs{1}_{-6,-2}+ \bfs{5}_{0,0}+\bfs{3}_{0,0}+\bfs{3}_{4,0}+\bfs{3}_{-4,0}\\
 &\qquad\qquad\qquad\qquad\qquad\qquad\qquad\qquad+\bfs{1}_{0,0}+ \bfs{5}_{-2,2}+\bfs{1}_{-2,2}+\bfs{3}_{ 2, 2}+\bfs{1}_{6, 2}\,.
\end{split}
\end{equation}
To determine the commutant of $\so(3)$ in $\frak e_{7(7)}$ in \eqref {embed}, we also need
\begin{equation}\label{}
\begin{split}
{\bf 35}_v \quad & \longrightarrow\quad {\bf 10}_2+{\bf 15}_0+\overline{\bf 10}_{-2}\\
& \longrightarrow \quad \overline{\bf 6}_{2,2}+{\bf 3}_{-2,2}+{\bf 1}_{-6,2}+{\bf 8}_{0,0}+{\bf 3}_{4,0}+\overline{\bf 3}_{-4,0}+{\bf 1}_{0,0}+{\bf 6}_{-2,-2}+\overline{\bf 3}_{2,-2}+{\bf 1}_{6,-2}\\
& \longrightarrow \quad \bfs 5_{2,2}+\bfs 1_{2,2}+\bfs 3_{-2,2}+{\bf 1}_{-6,2}+\bfs 5_{0,0}+\bfs 3_{0,0}+\bfs 3_{4,0} +\bfs 3_{-4,0}+\bfs 1_{0,0}\\ & \hspace{3.22 in}+\bfs 5_{-2,-2}+\bfs 1_{-2,-2} +\bfs 3_{2,-2}+\bfs 1_{6,-2}\,.
\end{split}
\end{equation}

Finally for the spin-1/2 fields we have
\begin{equation}\label{eq:56vbranch}
\begin{split}
\bfs{56}_v\quad\longrightarrow\quad & \overline{\bfs{20}}_{-1} + \bfs{20}_{1}+\bfs{4}_{1}+\bar{\bfs{4}}_{-1}+\bfs{4}_{-3}+\bar{\bfs{4}}_{3}\\
\quad\longrightarrow\quad & \bar{\bfs 6}_{-1,-1}+\bfs{8}_{3,-1}+\bfs{3}_{-5,-1}+\bar{\bfs{3}}_{-1,-1}+ \bfs{6}_{1,1}+\bfs{8}_{-3,1}+\bar{\bfs{3}}_{5,1}+\bfs{3}_{1,1}\\
&+\bfs{3}_{1,1}+\bfs{1}_{-3,1}+\bar{\bfs{3}}_{-1,-1}+\bfs{1}_{3,-1}+\bfs{3}_{1,-3}+\bfs{1}_{-3,-3}+\bar{\bfs{3}}_{-1,3}+\bfs{1}_{3,3}\\
\quad\longrightarrow\quad & \bfs{5}_{-1,-1}+\bfs{1}_{-1,-1}+\bfs{5}_{3,-1}+\bfs{3}_{3,-1}+\bfs{3}_{-5,-1}+\bfs{3}_{-1,-1}+ \bfs{5}_{1,1}+ \bfs{1}_{1,1}+\bfs{5}_{-3,1}+\bfs{3}_{-3,1}\\&+\bfs{3}_{5,1}+\bfs{3}_{1,1}+\bfs{3}_{1,1}+\bfs{1}_{-3,1}+\bfs{3}_{-1,-1}+\bfs{1}_{3,-1}+\bfs{3}_{1,-3}+\bfs{1}_{-3,-3}+\bfs{3}_{-1,3}+\bfs{1}_{3,3}\,.
\end{split}
\end{equation}
The singlets under $\so(3)$ are the metric, two Abelian gauge fields, 2 spin-3/2 fields, 10 scalars, and 6 spin-1/2 fields. This is precisely the matter contents of a four-dimensional $\mathcal{N}=2$ gauged supergravity coupled to 1 vector multiplet and 2 full hypermultiplets.

%%%%%%%%%%%%%%%%%%%%
\section{The full spectrum of $\mathcal{N}=8$ supergravity}
\label{AppendixB}
%%%%%%%%%%%%%%%%%%%%

In this appendix we present the masses of all bosonic and fermionic fields of the four-dimensional $\mathcal{N}=8$ supergravity around the $\SO(3)$ invariant $\mathcal{N}=1$ AdS$_4$ vacuum, $\tt S1384096$, studied in the main text. The spectrum of the spin-0, spin-1/2, and spin-3/2 fields and their $\SO(3)$ representations were already presented in \cite{Comsa:2019rcz} and is summarized in Table~\ref{tbl-m0}, Table~\ref{tbl-m12}, and Table~\ref{tbl-m32}, respectively. The spin-2 graviton is of course massless and not charged under $\SO(3)$ and the spectrum and $\SO(3)$ representations of the spin-1 vector fields are presented in Table~\ref{tbl-m1}. The latter  masses are computed using the general mass formulae for spin-1 fields in \cite{Trigiante:2016mnt}.

Using the standard AdS/CFT dictionary, see \cite{DHoker:2002nbb} for a review, the spectrum of these fields is mapped to the spectrum of operators in the dual $\mathcal{N}=1$ SCFT. The conformal dimensions of the dual operators of spin $s$ can be computed using formulae in Table~\ref{tbl-dims}. We use the dimensionless mass $mL$ of the supergravity fields, where $L$ is the AdS$_4$ scale. We have also indicated a reference where the derivation of each of the formulae can be found.

To understand how the spectrum of supergravity excitations maps to operators in the dual three-dimensional $\mathcal{N}=1$ SCFT it is useful to recall some aspects of the representation theory of the $\mathcal{N}=1$ superconformal algebra, see \cite{Cordova:2016emh} for a recent discussion. Operators in the $\mathcal{N}=1$ SCFT are labelled by their conformal dimension $\Delta$ and spin $s$ and will be denoted by $|\Delta,s\rangle$.\footnote{The authors of \cite{Cordova:2016emh} label the operators with $j=2s$.} These operators belong to one of the superconformal multiplets summarized in Table~\ref{tbl-N1rep}.

Using the information in Tables~\ref{tbl-m0}-\ref{tbl-m32} we can organize the spectrum of operators in the $\mathcal{N}=1$ SCFT dual to the $\SO(3)$ AdS$_4$ vacuum in the following superconformal multiplets:

%%%%%%%%%%%%%%%%%%%%%%%%%%%%%%%%%%%%%%%%%%%%%%%%%%
\begin{table}[H]
\renewcommand{\arraystretch}{1.0}
\begin{center}
\begin{tabular}{@{\extracolsep{20 pt}}l c c c }
\toprule
\noalign{\smallskip}
$\#$ \hspace{-10 pt} &  $m^2L^2$ & $\SO(3)$ irreps & $\Delta$ \\
\noalign{\smallskip}
\midrule
\noalign{\smallskip}
1$*$ & $16.26186$ & $\bf 1$ & $5.802541$    \\
\noalign{\smallskip}
1 & $16.09544$ & $\bf 1 $ & $5.783158$    \\
\noalign{\smallskip}
1$*$ & $8.656777$ & $\bf 1$ & $4.802541$    \\
\noalign{\smallskip}
1 & $8.529126$ & $\bf 1$ & $4.783158$ \\
\noalign{\smallskip}
1$*$ & $8.094691$ & $\bf 1 $ & $4.716316$ \\
\noalign{\smallskip}
3 & $ 5.322114$ & $\bf 3$ & $4.251747$ \\
\noalign{\smallskip}
3 & $ 5.182218$ & $\bf 3$ & $4.226210$ \\
\noalign{\smallskip}
5 & $ 3.817573$ & $\bf 5$ & $3.963244$ \\
\noalign{\smallskip}
1$*$ & $ 2.662058$ & $\bf 1$ & $3.716316$ \\
\noalign{\smallskip}
25 & $0$ & ${\bf 5}\oplus 6\times{\bf 3}\oplus 2\times {\bf 1} $ & $3$ \\
\noalign{\smallskip}
5 & $-0.108916$ & $\bf{5} $ & $2.963244$ \\
\noalign{\smallskip}
5 & $-1.099493$ & $\bf{5} $ & $2.572617$ \\
\noalign{\smallskip}
8 & $-1.396494$ & ${\bf 5}\oplus {\bf 3} $ & $0.5761462$ or $2.423854$ \\
\noalign{\smallskip}
1$*$ & $-1.685601$ & ${\bf1} $ & $0.7487353$ or $2.251265$ \\
\noalign{\smallskip}
1$*$ & $-2.188131$ & ${\bf 1} $ & $1.251265$ or $1.748735$ \\
\noalign{\smallskip}
3 & $-2.244202$ & ${\bf 3} $ & $1.423854$ or $1.576146$ \\
\noalign{\smallskip}
5 & $-2.244727$ & ${\bf 5} $ & $1.427383$ or $1.572617$ \\
\noalign{\smallskip}
\bottomrule
\end{tabular}
\caption{
Masses of the 70 supergravity scalars at the $\mathcal{N}=1$ $\SO(3)$-invariant point, the corresponding $\SO(3)$ representations, and the conformal dimensions of the dual operators. The conformal dimensions are obtained using the standard AdS/CFT formula $m^2L^2 = \Delta(\Delta-3)$ and choosing the root of this quadratic equation which obeys the unitarity bound $\Delta\geq1/2$. When $-9/4\leq m^2L^2 <5/4$ one has a choice of alternate quantization, see \cite{Klebanov:1999tb}, and both possible conformal dimensions are presented.}
\label{tbl-m0}
\end{center}
\end{table}
%%%%%%%%%%%%%%%%%%%%%%%%%%%%%%%%%%%%%%%%%%%%%%%%%%

\begin{itemize}[leftmargin=*]
\item \textit{Short spin-3/2:} \quad  This is simply the energy momentum multiplet which is neutral under $\mathfrak{so}(3)$ and contains the following operators
\begin{equation}
|\tfrac{5}{2},\tfrac{3}{2}\rangle\, \qquad\qquad |3,2\rangle\,.
\end{equation}
The supergravity modes corresponding to these operators are the spin-3/2 mode in the last line of Table~\ref{tbl-m32} and the metric.

%
%%%%%%%%%%%%%%%%%%%%%%%%%%%%%%%%%%%%%%%%%%%%%%%%%%
\begin{table}[H]
\renewcommand{\arraystretch}{1.0}
\begin{center}
\begin{tabular}{@{\extracolsep{25 pt}}r c c c }
\toprule
\noalign{\smallskip}
$\#$ \hspace{-10 pt} & $m^2L^2$ & $\SO(3)$ irreps & $\Delta$ \\
\noalign{\smallskip}
\midrule
\noalign{\smallskip}
1 & $14.45932$ & $\bf 1$ & $5.302541$    \\
1 & $14.31228$ & $\bf 1 $ & $5.283158$    \\
3 & $13.64583$ & $\bf 3$ & $5.194026$    \\
1 & $12.71861$ & $\bf 1$ & $5.066316$ \\
3 & $11.62481$ & $\bf 3 $ & $4.909518$ \\
3 & $10.57386$ & $\bf 3$ & $4.751747$ \\
3 & $ 10.40843$ & $\bf 3$ & $4.726210$ \\
1 & $ 7.378374$ & $\bf 1$ & $4.216316$ \\
3 & $ 5.070367$ & $\bf 3$ & $3.751747$ \\
3 & $4.956009$ & ${\bf 3}$ & $3.726210$ \\
5 & $3.854328$ & $\bf{5} $ & $3.463244$ \\
3 & $3.411457$ & $\bf{3} $ & $3.347013$ \\
1 & $3.179652$ & ${\bf 1} $ & $3.283158$ \\
3 & $2.906203$ & ${\bf3} $ &  $3.204759$ \\
5 & $2.027360$ & ${\bf 5} $ &  $2.923854$ \\
5 & $0.3278901$ & ${\bf 5} $ & $2.072617$ \\
8 & $0.1796520$ & ${\bf 5}\oplus{\bf 3} $ &  $1.923854$ \\
1 & $0.0631340$ & ${\bf 1} $ &  $1.751265$ \\
3 & $0$ & ${\bf 3} $ &  $3/2$ \\
\noalign{\smallskip}
\bottomrule
\end{tabular}
\caption{Masses of the 56 spin-1/2 supergravity fermions at the $\mathcal{N}=1$ $\SO(3)$-invariant point, the corresponding $\SO(3)$ representations, and the conformal dimensions of the dual operators.}
\label{tbl-m12}
\end{center}
\end{table}
%%%%%%%%%%%%%%%%%%%%%%%%%%%%%%%%%%%%%%%%%%%%%%%%%%

\item \textit{Short spin-1/2:}\quad  This is the conserved $\mathfrak{so}(3)$ current and contains the following operators
\begin{equation}
|\tfrac{3}{2},\tfrac{1}{2}\rangle\, \qquad\qquad |2,1\rangle\,.
\end{equation}
The supergravity modes corresponding to these operators are in the $\mathbf{3}$ of $\mathfrak{so}(3)$ and correspond to the massless vector and spin-1/2 modes in Table~\ref{tbl-m1} and Table~\ref{tbl-m12}, respectively.

\item \textit{Long spin-1:}\quad  There are three such multiplets. We present all operators in them below and indicate also the  $\mathfrak{so}(3)$ representations
\begin{equation}
\begin{split}
&|2.704760,1\rangle\, \qquad |3.204759,\tfrac{1}{2}\rangle \qquad |3.204759,\tfrac{3}{2}\rangle \qquad |3.704760,1\rangle\,, \qquad \mathbf{3}\,,\\
&|2.783158,1\rangle\, \qquad |3.283158,\tfrac{1}{2}\rangle \qquad |3.283158,\tfrac{3}{2}\rangle \qquad |3.783158,1\rangle\,, \qquad \mathbf{1}\,,\\
&|2.847013,1\rangle\, \qquad |3.347013,\tfrac{1}{2}\rangle \qquad |3.347013,\tfrac{3}{2}\rangle \qquad |3.847013,1\rangle\,, \qquad \mathbf{3}\,.
\end{split}
\end{equation}
The supergravity modes corresponding to these operators are the spin-$\frac{1}{2}$, spin-$1$, and spin-$\frac{3}{2}$ modes of the corresponding dimension in Table~\ref{tbl-m12}, Table~\ref{tbl-m1}, and Table~\ref{tbl-m32}, respectively.

%
%%%%%%%%%%%%%%%%%%%%%%%%%%%%%%%%%%%%%%%%%%%%%%%%%%
\begin{table}[t]
\renewcommand{\arraystretch}{1.0}
\begin{center}
\begin{tabular}{@{\extracolsep{15 pt}}r c c c }
\toprule
\noalign{\smallskip}
$\#$ \hspace{-10 pt} & $m^2L^2$ & $\SO(3)$ irreps & $\Delta$ \\
\noalign{\smallskip}
\midrule
3 & $7.322116$ & $\bf 3$ & $4.251748$    \\
3 & $7.182220$ & $\bf 3 $ & $4.226210$    \\
3 & $5.258471$ & $\bf 3$ & $3.847013$    \\
1 & $4.962811$ & $\bf 1$ & $3.783158$ \\
3 & $4.610963$ & $\bf 3 $ & $3.704760$ \\
3 & $ 1.564444$ & $\bf 3$ & $2.847013$ \\
1 & $ 1.396495$ & $\bf 1$ & $2.783158$ \\
3 & $ 1.201444$ & $\bf 3$ & $2.704760$ \\
5 & $ 0.603506$ & $\bf 5$ & $2.423854$ \\
3 & $0$ & ${\bf 3}$ & $2$ \\
\noalign{\smallskip}
\bottomrule
\end{tabular}
\caption{Masses of the 28 supergravity vector fields at the $\mathcal{N}=1$ $\SO(3)$-invariant point, the corresponding $\SO(3)$ representations, and the conformal dimensions of the dual operators.}
\label{tbl-m1}
\end{center}
\end{table}
%%%%%%%%%%%%%%%%%%%%%%%%%%%%%%%%%%%%%%%%%%%%%%%%%%

%
%%%%%%%%%%%%%%%%%%%%%%%%%%%%%%%%%%%%%%%%%%%%%%%%%%
\begin{table}[t]
\renewcommand{\arraystretch}{1.0}
\begin{center}
\begin{tabular}{@{\extracolsep{25 pt}}l c c c }
\toprule
\noalign{\smallskip}
$\#$ \hspace{-10 pt} & $m^2L^2$ & $\SO(3)$ irreps & $\Delta$ \\
\noalign{\smallskip}
\midrule
\noalign{\smallskip}
3 & $3.411457$ & $\bf 3$ & $3.347013$    \\
1 & $3.179652$ & $\bf 1 $ & $3.283158$    \\
3 & $2.906203$ & $\bf 3$ & $3.204759$    \\
1$*$ & $1$ & $\bf 1$ & $5/2$ \\
\noalign{\smallskip}
\bottomrule
\end{tabular}
\caption{Masses of the 8 spin-3/2 supergravity fermions at the $\mathcal{N}=1$ $\SO(3)$-invariant point, the corresponding $\SO(3)$ representations, and the conformal dimensions of the dual operators. }
\label{tbl-m32}
\end{center}
\end{table}
%%%%%%%%%%%%%%%%%%%%%%%%%%%%%%%%%%%%%%%%%%%%%%%%%%

%%%%%%%%%%%%%%%%%%%%%%%%%%%%%%%%%%%%%%%%%%%%%%%%%%
\begin{table}[t]
\renewcommand{\arraystretch}{1.4}
\begin{center}
\begin{tabular}{@{\extracolsep{15 pt}}l l c }
\toprule
  Spin & Dimension &   \\
\midrule
\noalign{\smallskip}
0 & $\Delta =\frac{3}{2}\pm \sqrt{\frac{9}{4}+m^2L^2}$ & \cite{DHoker:2002nbb}\\
${1\over 2}$ & $\Delta  =\frac{3}{2}+|mL|$ & \cite{Henningson:1998cd}\\
$1$ & $\Delta =\frac{3}{2}\pm \sqrt{\frac{1}{4}+m^2L^2}$ & \cite{lYi:1998trg} \\
${3\over 2}$ & $\Delta =\frac{3}{2}+|mL|$ & \cite{Volovich:1998tj,Corley:1998qg}\\
\bottomrule
\end{tabular}
\caption{Dimensions of operators dual to fields of spin, $s$, and mass, $m$. }
\label{tbl-dims}
\end{center}
\end{table}
%%%%%%%%%%%%%%%%%%%%%%%%%%%%%%%%%%%%%%%%%%%%%%%%%%

%%%%%%%%%%%%%%%%%%%%%%%%%%%%%%%%%%%%%%%%%%%%%%%%%%
\begin{table}[t]
\renewcommand{\arraystretch}{1.2}
\begin{center}
\begin{tabular}{@{\extracolsep{2 pt}}l c c c }
\toprule
\noalign{\smallskip}
Name \hspace{-10 pt} &  Primary & Descendants & Unitarity bound \\
\noalign{\smallskip}
\midrule
\noalign{\smallskip}
Identity ($B_1$) & $|0,0\rangle$ & - & $\Delta=0$    \\
Short scalar  ($A_2'$)& $|\frac{1}{2},0\rangle$ & $|\frac{1}{2},\frac{1}{2}\rangle$ & $\Delta=\frac{1}{2}$    \\
Short spin  ($A_1$)& $|s+1,s\rangle$ & $|s+\frac{3}{2},s+\frac{1}{2}\rangle$ & $\Delta=s+1$; $s>0$    \\
Long scalar  ($L'$ )& $|\Delta,0\rangle$ & $|\Delta+\frac{1}{2},\frac{1}{2}\rangle$; $|\Delta+1,0\rangle$ & $\Delta>\frac{1}{2}$    \\
Long spin  ($L$)& $|\Delta,s\rangle$ & $|\Delta+\frac{1}{2},s+\frac{1}{2}\rangle$; $|\Delta+\frac{1}{2},s-\frac{1}{2}\rangle$; $|\Delta+1,s\rangle$ & $\Delta>s+1$; $s>0$    \\
\bottomrule
\end{tabular}
\caption{The $\mathcal{N}=1$ superconformal multiplets. The first column indicates also the notation for each multiplet used in \cite{Cordova:2016emh}.}
\label{tbl-N1rep}
\end{center}
\end{table}
%%%%%%%%%%%%%%%%%%%%%%%%%%%%%%%%%%%%%%%%%%%%%%%%%%

\item \textit{Long spin-$1/2$:}\quad  There are three such multiplets. We present all operators in them below and indicate also the  $\mathfrak{so}(3)$ representations
\begin{equation}
\begin{split}
&|3.751747,\tfrac{1}{2}\rangle\, \qquad |4.251748,1\rangle \qquad |4.251747,0\rangle \qquad |4.751747,\tfrac{1}{2}\rangle\,, \qquad \mathbf{3}\,,\\
&|3.726210,\tfrac{1}{2}\rangle\, \qquad |4.226210,1\rangle \qquad |4.226210,0\rangle \qquad |4.726210,\tfrac{1}{2}\rangle\,, \qquad \mathbf{3}\,,\\
&|1.923854,\tfrac{1}{2}\rangle\, \qquad |2.423854,1\rangle \qquad |2.423854,0\rangle \qquad |2.923854,\tfrac{1}{2}\rangle\,, \qquad \mathbf{5}\,.
\end{split}
\end{equation}
The supergravity modes corresponding to these operators are the spin-0, spin-$\frac{1}{2}$, and spin-$1$ modes of the corresponding dimension in Table~\ref{tbl-m0}, Table~\ref{tbl-m12}, and Table~\ref{tbl-m1}, respectively.

\item \textit{Long scalar:}\quad  There are seven such multiplets. We present all operators in them along with their  $\mathfrak{so}(3)$ representations below
\begin{equation}
\begin{split}
&|4.802541,0\rangle\, \qquad |5.302541,\tfrac{1}{2}\rangle \qquad |5.802541,0\rangle\,, \qquad \mathbf{1}\,,\\
&|4.783158,0\rangle\, \qquad |5.283158,\tfrac{1}{2}\rangle \qquad |5.783158,0\rangle\,, \qquad \mathbf{1}\,,\\
&|3.716316,0\rangle\, \qquad |4.216316,\tfrac{1}{2}\rangle \qquad |4.716316,0\rangle\,, \qquad \mathbf{1}\,,\\
&|2.963244,0\rangle\, \qquad |3.463244,\tfrac{1}{2}\rangle \qquad |3.963244,0\rangle\,, \qquad \mathbf{5}\,,\\
&|1.572617,0\rangle\, \qquad |2.072617,\tfrac{1}{2}\rangle \qquad |2.572617,0\rangle\,, \qquad \mathbf{5}\,,\\
&|1.423854,0\rangle\, \qquad |1.923854,\tfrac{1}{2}\rangle \qquad |2.423854,0\rangle\,, \qquad \mathbf{3}\,,\\
&|1.251265,0\rangle\, \qquad |1.751265,\tfrac{1}{2}\rangle \qquad |2.251265,0\rangle\,, \qquad \mathbf{1}\,.\\
\end{split}
\end{equation}
The supergravity modes corresponding to these operators are the spin-0 and spin-$\frac{1}{2}$ modes of the corresponding dimension in Table~\ref{tbl-m0} and Table~\ref{tbl-m12}, respectively.

\end{itemize}

In addition to the modes discussed above it should be noted that due to the spontaneous breaking of the $\mathcal{N}=8$ supersymmetry and the $\mathfrak{so}(8)$ gauge symmetry of the supergravity theory there are spin-0 and spin-$1/2$ modes that are ``eaten'' by the usual (super)Higgs mechanism. These are the spin-1/2 modes in the 3rd, 4th and 5th line of Table~\ref{tbl-m12} as well as the spin-0 mode in the 10th line of Table~\ref{tbl-m0}.

It should also be noted that for spin-0 modes with mass in the range $-\frac{9}{4} \leq m^2L^2<-\frac{5}{4}$ there is an ambiguity in assigning a conformal dimension of the dual CFT operator. This happens because for both choices of sign in Table~\ref{tbl-dims} $\Delta$ obeys the unitarity bound. Invoking supersymmetry however uniquely fixes the choice of sign in Table~\ref{tbl-dims} and we have chosen the only possible sign that allows for organizing the bottom five entries in Table~\ref{tbl-m0} into $\mathcal{N}=1$ superconformal multiplets.

%
%%%%%%%%%%%%%%%%%%%%
\section{Discrete SO(8) rotations}
\label{AppendixC}
%%%%%%%%%%%%%%%%%%%%

Consider the $\SO(3)$  subgroup of $\SO(8)$ introduced in Section~\ref{sec:truncation}. It is straightforward to check that any  $\rm SO(8)$ rotation that commutes with this subgroup must be of the form 
\begin{equation}\label{comrot}
  \left(\begin{matrix}
\cos \alpha  & 0 & 0 & \sin \alpha  & 0 & 0 & 0 & 0 \\
 0 & \cos \alpha  & 0 & 0 & \sin \alpha  & 0 & 0 & 0 \\
 0 & 0 & \cos \alpha  & 0 & 0 & \sin \alpha  & 0 & 0 \\
 \mp\sin \alpha  & 0 & 0 & \pm\cos \alpha  & 0 & 0 & 0 & 0 \\
 0 & \mp\sin \alpha  & 0 & 0 & \pm\cos \alpha  & 0 & 0 & 0 \\
 0 & 0 & \mp\sin \alpha  & 0 & 0 & \pm\cos \alpha  & 0 & 0 \\
 0 & 0 & 0 & 0 & 0 & 0 & \cos \beta & \sin \beta \\
 0 & 0 & 0 & 0 & 0 & 0 & \mp\sin \beta & \pm\cos \beta  
\end{matrix}\right)\,.
\end{equation}

Choosing the upper sign, we obtain the two parameter family of rotations, $g(\alpha,\beta)$, corresponding to the  $\rU(1)\times \rU(1)$ gauge group of the $\cals N=2$ supergravity of the $\SO(3)$-invariant truncation. For the lower sign, the rotations  are given by 
\begin{equation}\label{}
g(-\alpha,-\beta+\pi)\,g_S\,,
\end{equation}
where $g_S$ is the generator \eqref{gS} of the discrete $\ZZ_2$ symmetry that defines our truncation. 

Since the scalar fields $s_a$, $a=1,\ldots,6$, are by construction invariant under $g_S$, any residual nontrivial action of the $\SO(8)$ subgroup given in  \eqref{comrot} on the coset  \eqref{eq:su11coset} must come from the $\rU(1)\times \rU(1)$ transformations that preserve  \eqref {ttrunc}. 
At a generic point in the coset, this leaves a discrete $\ZZ_2\times \ZZ_4$ subgroup of $\SO(8)$ generated by $g_H\equiv g(0,\pi)$  and $g_C\equiv g(\pi/2,\pi/2)$, which are elements of order 2 and 4, respectively. However, since $g_C^2\eql g(\pi,\pi)$ acts trivially on the coset, we end up with only $\ZZ_2\times \ZZ_2$ worth of  $\SO(8)$ rotations that preserve the scalar manifold in the $\SO(3)\times \ZZ_2$-invariant truncation. Those are generated by the transformations
\begin{equation}\label{dtrans}
\begin{split}
g_H & ~:~(z_1,z_2,z_3)~\longrightarrow  ~(-z_1,-z_2,z_3) \,,\\
g_C& ~:~(z_1,z_2,z_3)~\longrightarrow~(-\bar z_1,-\bar z_2,z_3)\,.
\end{split}
\end{equation}
Hence any point in the coset lies on an $\SO(8)$ orbit  obtained by acting on that point with $1$, $g_H$, $g_C$ and $g_Hg_C$. When both $z_1$ and $z_2$ are either real or imaginary, the orbit degenerates to two points.

At special points of the coset there might be additional rotations, $g(\alpha,\beta)$, that map it onto another point on the coset. In particular, this happens for two critical points, $\S0880733 $ and $\S1039230$, in Section~\ref{sec:otherpoints}, with the special rotations given by $g_R\equiv g(-3\pi/4,\pi/4)$ and $g_R'\equiv g(-\pi/4,\pi/4)$, respectively.

While both rotations in \eqref{dtrans} are obviously symmetries of the potential \eqref{eq:cPdef}, only $g_H$  preserves the superpotential \eqref{eq:cWdef}. This is just the reflection of the fact that the second transformation acts nontrivially on the supersymmetry by mapping the $\cals N=1$ supergravity given by the $\SO(3)\times \ZZ_2$-invariant truncation to an equivalent one obtained by a different discrete $\ZZ_2$ symmetry.

Finally, we note that 4-form $\Phi^{(3)}$ in \eqref{critsub}   is invariant under the transformations \eqref{comrot}. In particular, this implies  that $z_3$ is invariant under the discrete symmetries above.

%%%%%%%%%%%%%%%%%%%%
\section{New critical points S2096313 and S2443607}
\label{appNew}
%%%%%%%%%%%%%%%%%%%%
In this appendix we present   numerical data for the two new critical points,  $\S2096313$ and $\S2443607$, in the same  format as in \cite{Comsa:2019rcz}. The parametrization of the scalar coset \eqref{eq:su11coset}  in terms of $s_a$, $a=1,\ldots,6$ introduced  in Section~\ref{sec:truncation}  is related to  the one in  \cite{Comsa:2019rcz} and in the tables below by: 
\begin{align}\notag
{A} &\eql  \frac{1}{8} \left(2 s_1+s_5\right), &  {B} &\eql \frac{1}{8} \left(s_1-s_3+2
   s_5\right), & {C} &\eql  \frac{1}{8} \left(s_5-2 s_1\right),\\   {D} &\eql  \frac{1}{8} \left(s_1-s_3-2
   s_5\right), & {E} &\eql  -\frac{s_5}{8}, & {F}& \eql  \frac{1}{8} \left(-s_1-s_3\right), \label{changvar}\\ \notag{G}& \eql 
   \frac{1}{8} \left(s_2+s_4-s_6\right), &  {H}& \eql  \frac{1}{8} \left(-s_2-s_4-s_6\right),&  {I}& \eql 
   \frac{1}{8} \left(-3 s_2+s_4+3 s_6\right), \\ {J}& \eql  \frac{1}{8} \left(3 s_2-s_4+3 s_6\right)\,,\notag
\end{align}
where the parameters $A,\ldots,J$ satisfy
\begin{equation}\label{}
A-B-E+F=0\,,\quad  B-D+4E=0\,,\quad  C+D-E-F=0\,,\quad 3G+3H+I+J=0\,.
\end{equation}

\newpage

\subsection{Point S2096313}
\label{S:S2096313}
{\small\begin{longtable}{L}
{\bf S2096313}: \mathfrak{so}(8)\to {\mathfrak{so}(3)}+{\mathfrak{u}(1)}\\
V/g^2\approx-20.9631372891\\
\begin{minipage}[t]{12cm}\begin{flushleft}$\mathbf{8}_{c}\to2\times\mathbf{3} + \mathbf{1}^{\scriptscriptstyle ++} + \mathbf{1}^{\scriptscriptstyle --},\;$ $\mathbf{8}_{v,s}\to\mathbf{3}^{\scriptscriptstyle +} + \mathbf{3}^{\scriptscriptstyle -} + \mathbf{1}^{\scriptscriptstyle +} + \mathbf{1}^{\scriptscriptstyle -},\;$ $\mathbf{28}_{}\to\mathbf{5} + 2\times\mathbf{3}^{\scriptscriptstyle ++} + 3\times\mathbf{3} + 2\times\mathbf{3}^{\scriptscriptstyle --} + 2\times\mathbf{1}$\end{flushleft}\end{minipage}\\
m^2/m_0^2[\psi]:\begin{minipage}[t]{10cm}\begin{flushleft}$7.750^{}_{\mathbf{1}^{\scriptscriptstyle +} + \mathbf{1}^{\scriptscriptstyle -}}$, $2.550^{}_{\mathbf{3}^{\scriptscriptstyle +} + \mathbf{3}^{\scriptscriptstyle -}}$\end{flushleft}\end{minipage}\\
m^2/m_0^2[\chi]:\begin{minipage}[t]{10cm}\begin{flushleft}$15.500_{\mathbf{1}^{\scriptscriptstyle +} + \mathbf{1}^{\scriptscriptstyle -}}$, $7.875_{\mathbf{3}^{\scriptscriptstyle +++} + \mathbf{3}^{\scriptscriptstyle ---}}$, $5.137_{\mathbf{3}^{\scriptscriptstyle +} + \mathbf{3}^{\scriptscriptstyle -}}$, $5.100_{\mathbf{3}^{\scriptscriptstyle +} + \mathbf{3}^{\scriptscriptstyle -}}$, $4.596_{\mathbf{5}^{\scriptscriptstyle +} + \mathbf{5}^{\scriptscriptstyle -}}$, $2.475_{\mathbf{1}^{\scriptscriptstyle +++} + \mathbf{1}^{\scriptscriptstyle ---}}$, $0.875_{\mathbf{1}^{\scriptscriptstyle +} + \mathbf{1}^{\scriptscriptstyle -}}$, $0.754_{\mathbf{5}^{\scriptscriptstyle +} + \mathbf{5}^{\scriptscriptstyle -}}$, $0.727_{\mathbf{3}^{\scriptscriptstyle +} + \mathbf{3}^{\scriptscriptstyle -}}$, $0.160_{\mathbf{3}^{\scriptscriptstyle +} + \mathbf{3}^{\scriptscriptstyle -}}$\end{flushleft}\end{minipage}\\
m^2/m_0^2[\phi]:\begin{minipage}[t]{10cm}\begin{flushleft}$18.400^{c}_{\mathbf{1}^{\scriptscriptstyle ++++} + \mathbf{1}^{\scriptscriptstyle ----}}$, $10.000^{s}_{\mathbf{1}^{\scriptscriptstyle ++} + \mathbf{1}^{\scriptscriptstyle --}}$, $9.600^{s}_{\mathbf{5}^{\scriptscriptstyle ++} + \mathbf{5}^{\scriptscriptstyle --}}$, $8.728^{m}_{\mathbf{1}}$, $8.596^{m}_{\mathbf{5}}$, $0.400^{m}_{\mathbf{3}}$, $0.000^{s}_{\mathbf{3}^{\scriptscriptstyle ++} + \mathbf{3} + \mathbf{3}^{\scriptscriptstyle --}}$, $0.000^{c}_{\mathbf{5} + 2\times\mathbf{3}^{\scriptscriptstyle ++} + 2\times\mathbf{3}^{\scriptscriptstyle --} + \mathbf{1}}$, $0.000^{m}_{\mathbf{3}}$, $-0.596^{m}_{\mathbf{5}}$, $-0.612^{m}_{\mathbf{1}}$, $-1.200^{s}_{\mathbf{1}^{\scriptscriptstyle ++} + \mathbf{1}^{\scriptscriptstyle --}}$, $-2.000^{m}_{\mathbf{3}}$, $-2.400^{m*}_{\mathbf{5}}$, $-2.516^{m*}_{\mathbf{1}}$\end{flushleft}\end{minipage}\\[6 pt]
M_{\alpha\beta}=\rm{diag}\left(-3A, -3A, A, A, A, A, A, A\right)
\\
M_{\dot\alpha\dot\beta}=\rm{diag}\left(C, D, B, B, C, D, D, C\right)
\\
\begin{minipage}[t]{12cm}\begin{flushleft}
$A\approx-0.1641598793,\;$
$B\approx-0.5046414602,\;$
$C\approx-0.0723920925,\;$
$D\approx0.4088197326$
\end{flushleft}\end{minipage}
\end{longtable}}

\subsection{Point S2443607}
\label{S:S2443607}
{\small\begin{longtable}{L}
{\bf S2443607}: \mathfrak{so}(8)\to {\mathfrak{so}(3)}\\
V/g^2\approx-24.4360747652\\
\begin{minipage}[t]{12cm}\begin{flushleft}$\mathbf{8}_{v,s,c}\to2\times\mathbf{3} + 2\times\mathbf{1},\;$ $\mathbf{28}_{}\to\mathbf{5} + 7\times\mathbf{3} + 2\times\mathbf{1}$\end{flushleft}\end{minipage}\\
m^2/m_0^2[\psi]:\begin{minipage}[t]{10cm}\begin{flushleft}$11.568^{}_{\mathbf{1}}$, $11.463^{}_{\mathbf{1}}$, $4.354^{}_{\mathbf{3}}$, $3.726^{}_{\mathbf{3}}$\end{flushleft}\end{minipage}\\
m^2/m_0^2[\chi]:\begin{minipage}[t]{10cm}\begin{flushleft}$23.137_{\mathbf{1}}$, $22.926_{\mathbf{1}}$, $18.908_{\mathbf{3}}$, $18.908_{\mathbf{3}}$, $10.714_{\mathbf{1}}$, $10.711_{\mathbf{1}}$, $8.708_{\mathbf{3}}$, $7.451_{\mathbf{3}}$, $6.299_{\mathbf{3}}$, $6.141_{\mathbf{3}}$, $5.757_{\mathbf{5}}$, $5.744_{\mathbf{5}}$, $2.122_{\mathbf{3}}$, $2.064_{\mathbf{1}}$, $1.624_{\mathbf{5}}$, $1.417_{\mathbf{3}}$, $1.257_{\mathbf{5}}$, $1.255_{\mathbf{1}}$, $0.201_{\mathbf{3}}$, $0.082_{\mathbf{3}}$\end{flushleft}\end{minipage}\\
m^2/m_0^2[\phi]:\begin{minipage}[t]{10cm}\begin{flushleft}$55.474^{m}_{\mathbf{1}}$, $55.474^{m}_{\mathbf{1}}$, $20.600^{m}_{\mathbf{1}}$, $20.559^{m}_{\mathbf{1}}$, $19.887^{m}_{\mathbf{5}}$, $19.876^{m}_{\mathbf{5}}$, $8.040^{m}_{\mathbf{1}}$, $7.864^{m}_{\mathbf{3}}$, $7.464^{m}_{\mathbf{3}}$, $5.996^{m}_{\mathbf{5}}$, $4.125^{m}_{\mathbf{1}}$, $0.000^{s}_{\mathbf{3} + \mathbf{1}}$, $0.000^{m}_{\mathbf{5} + 5\times\mathbf{3} + \mathbf{1}}$, $-0.023^{m}_{\mathbf{3}}$, $-0.562^{m}_{\mathbf{1}}$, $-0.743^{m}_{\mathbf{5}}$, $-2.513^{m*}_{\mathbf{3}}$, $-2.836^{m*}_{\mathbf{5}}$, $-3.205^{m*}_{\mathbf{1}}$\end{flushleft}\vspace{1 pt}
\end{minipage}\\
M_{\alpha\beta}=\left(\begin{array}{rrrrrrrr}
A&\phantom-0&\phantom-0&\phantom-0&\phantom-0&B&\phantom-0&\phantom-0\\
\phantom-0&E&\phantom-0&\phantom-0&C&\phantom-0&\phantom-0&\phantom-0\\
\phantom-0&\phantom-0&F&\phantom-0&\phantom-0&\phantom-0&\phantom-0&D\\
\phantom-0&\phantom-0&\phantom-0&F&\phantom-0&\phantom-0&-D&\phantom-0\\
\phantom-0&C&\phantom-0&\phantom-0&E&\phantom-0&\phantom-0&\phantom-0\\
B&\phantom-0&\phantom-0&\phantom-0&\phantom-0&A&\phantom-0&\phantom-0\\
\phantom-0&\phantom-0&\phantom-0&-D&\phantom-0&\phantom-0&F&\phantom-0\\
\phantom-0&\phantom-0&D&\phantom-0&\phantom-0&\phantom-0&\phantom-0&F\\
\end{array}\right)\\
\\
M_{\dot\alpha\dot\beta}=\rm{diag}\left(G, G, G, H, H, H, I, J\right)
\\
\begin{minipage}[t]{12cm}\begin{flushleft}
$A\approx0.2540142811,\;$
$B\approx0.3965710947,\;$
$C\approx-0.1697655242,\;$
$D\approx0.0009726589,\;$
$E\approx0.0291540284,\;$
$F\approx-0.1415841547,\;$
$G\approx0.4106803872,\;$
$H\approx0.0401731985,\;$
$I\approx-0.6761535338,\;$
$J\approx-0.6764072234$
\end{flushleft}\end{minipage}
\end{longtable}}

%%%%%%%%%%%%%%%%%%%%
%%%%%%%%%%%%%
\subsection{Ancillary files}
\label{App:ansfiles}
%%%%%%%%%%%%%

Numerical data for the position of the two new critical points, $\tt S2096313$ and $\tt S2443607$, in the same format as in 
\cite{Comsa:2019rcz}, can be dowloaded from the arXiv repository at:

\citeurl{https://arxiv.org/src/1909.10969v1/anc/extrema/S2096313/location.py.txt}

\citeurl{https://arxiv.org/src/1909.10969v1/anc/extrema/S2443607/location.py.txt}\\[1ex]

\noindent Algebraic data on the locations of critical points $\S1424025$ and $\S2443607$, which give rise to formulae that are too complicated to be included in the text of this work, are available at:

\citeurl{https://arxiv.org/src/1909.10969v2/anc/extrema/S1424025/algebraic.py.txt}

\citeurl{https://arxiv.org/src/1909.10969v2/anc/extrema/S2443607/algebraic.py.txt}

%%%%%%%%%%%%%%%%%%%%%%%%
\section{Minimal polynomials}
\label{App:files}
%%%%%%%%%%%%%%%%%%%%%%%

The PSLQ algorithm used to obtain the minimal polynomials in Sections~\ref{sec:newpoint} and \ref{sec:otherpoints} may also be used directly in the complex domain.\footnote{We thank  Moritz Firsching for pointing this to us and computing all the minimal polynomials in this appendix.}

As our coordinates map the hyperbolic plane to the unit disk, the relevant minimal polynomials for coordinates will be palindromic. If~$\zeta$ is a zero, then~$\zeta^{-1}$ will also be a zero, and this invariance under~$\zeta\mapsto\zeta^{-1}$ makes the highest-order coefficient match the lowest-order coefficient, etc. It so turns out that for many critical points, a de-palindromizing substitution (via the inverse of a Zhukovsky transform,~$\zeta\mapsto \zeta + \zeta^{-1}$), followed by a re-scaling, again leads to a palindromic polynomial.

%%%%%%%%%%%%%
\subsection{Minimal polynomials for $\S1384096$}
\label{App:filespol}
%%%%%%%%%%%%%

The minimal polynomials for the coordinates of the new $\cals N=1$ vacuum, $\cals P_{z_i}(\zeta)$, $i=1,2,3$, turn out to be ``doubly'' palindromic, and this property can be exploited to simplify their presentation.
 
Let
\begin{equation}\label{}
\begin{split}
S_1(\zeta) & \eql \zeta ^{12}+2 \zeta ^{10}+387 \zeta ^8-7276 \zeta ^6+59179 \zeta ^4-248970 \zeta
   ^2+416025\,,\\[6 pt]
S_2(\zeta) & \eql 237169 \zeta ^{12}-5533444 \zeta ^{10}+54887568 \zeta ^8-295250296 \zeta ^6+905373664
   \zeta ^4 \\ & \quad -1496099520 \zeta ^2+1038128400\,,\\[6 pt]
S_3(\zeta) & \eql   3 \zeta ^6+30 \zeta ^5+191 \zeta ^4+690 \zeta ^3+1337 \zeta ^2+1314 \zeta +521\,.
\end{split}
\end{equation}
Define the polynomials, $M_i(\zeta)$, as follows
\begin{equation}\label{}
M_i(2 \zeta)\eql (4\zeta)^{\text{ord}(S_i)}S_i(\zeta+\zeta^{-1})\,,
\end{equation}
Then 
\begin{equation}\label{}
\cals P_{z_i}(\zeta)\eql \zeta^{\text{ord}(M_i)} M_i(\zeta+\zeta^{-1})\,,\qquad i=1,2,3\,,
\end{equation}
are the  minimal polynomials with integer coefficients for $\S1384096$.

\subsection{Minimal polynomials for $\S1424025$ and $\S2443607$}
\label{App:algebraicpositions}

The minimal polynomials for the coordinates of the critical points $\S1424025$ and $\S2443607$ also allow ``double depalindromization'', and this property was exploited to obtain these expressions. Unfortunately, they are too complicated to be shown in the text, having degrees 208, 208, and 52 for $\S1424025$, and degrees 464, 464, and 232 for $\S2443607$. The preprint of this article on arXiv.org provides executable Python code that lists and verifies these polynomials. This is available at:

\citeurl{https://arxiv.org/src/1909.10969v2/anc/extrema/S1424025/algebraic.py.txt}

\citeurl{https://arxiv.org/src/1909.10969v2/anc/extrema/S2443607/algebraic.py.txt}

%\newpage

\bibliography{4dN1-new}
\bibliographystyle{JHEP}

\end{document}